\documentclass[a4paper,twocolumn,preprint,article,11pt,bibnotes,accepted=2024-06-17]{quantumarticle}
\pdfoutput=1

\usepackage[utf8]{inputenc}
\usepackage[numbers]{natbib}
\usepackage{multirow}
\usepackage[T1]{fontenc}
\usepackage{graphicx}
\usepackage{dcolumn}
\usepackage{bm}
\usepackage{siunitx}
\usepackage[italicdiff]{physics}
\usepackage[colorlinks=true, allcolors=blue]{hyperref}
\usepackage[utf8]{inputenc}
\usepackage[toc,page,title]{appendix}
\usepackage{graphicx}
\usepackage{subcaption}
\usepackage{mwe}
\usepackage[T1]{fontenc}
\usepackage{tikz}
\usepackage{graphicx}
\usepackage{adjustbox}
\usetikzlibrary{arrows, hobby}
\usepackage{wrapfig}
\usepackage{graphicx}
\usepackage{amsmath}
\usepackage{amssymb}
\usepackage{hyperref}
\usepackage{ mathrsfs }
\usepackage{indentfirst}
\usepackage{braket}
\usepackage{color}
\usepackage{empheq}
\usepackage{lipsum,adjustbox}
\usetikzlibrary{shapes}
\usetikzlibrary{plotmarks}
\usepackage{tikz}	
\usetikzlibrary{backgrounds,fit,decorations.pathreplacing}  

\tikzset{pics/.cd,
collector/.style={code={
\draw[fill=gray!20] (0,0.5) arc(90:-90:0.75cm and 0.5cm) -- cycle;}},
splitter/.style={code={\draw[ultra thick] (#1:{sqrt(1/2)}) --
(#1+180:{sqrt(1/2)});}},splitter/.default=135}

\tikzstyle{beamsplitter}=[fill=blue, fill opacity=0.2]

\newcommand{\bmtx}[1]{\begin{bmatrix}#1\end{bmatrix}}
\DeclareMathOperator{\diag}{diag}
\DeclarePairedDelimiter{\mean}{\langle}{\rangle}

\newcommand{\hc}{^{\dag}} 
\newcommand{\ah}{\hat{a}}
\newcommand{\eh}{\hat{e}}
\newcommand{\bh}{\hat{b}}
\newcommand{\fh}{\hat{f}}

\newcommand{\mbf}[1]{\mathbf{#1}}
\newcommand{\bi}{\mbf{i}}
\newcommand{\bj}{\mbf{j}}
\newcommand{\bn}{\mbf{n}}
\newcommand{\bbeta}{\bm{\beta}}

\newcommand{\win}{\omega_{\bi,\bn}}
\newcommand{\outijn}{\ketbra{\win}{\omega_{\bj,\bn}}}
\newcommand{\I}{\mathbb{I}}

\begin{document}
\title{{Comparison of Discrete Variable and Continuous Variable Quantum Key Distribution Protocols with Phase Noise in the Thermal-Loss Channel}}

\author{S. P. Kish} \email{sebastian.kish@data61.csiro.au}
 \affiliation{Data61, CSIRO, Marsfield, NSW, Australia.}
  \affiliation{Centre of Excellence for Quantum Computation and Communication Technology, Department of Quantum Science and Technology, Research School of Physics, The Australian National University, Canberra, ACT, Australia.}
\author{P. J. Gleeson}
 \affiliation{Centre of Excellence for Quantum Computation and Communication Technology, Department of Quantum Science and Technology, Research School of Physics, The Australian National University, Canberra, ACT, Australia.}
 \author{A. Walsh}
 \affiliation{Centre of Excellence for Quantum Computation and Communication Technology, Department of Quantum Science and Technology, Research School of Physics, The Australian National University, Canberra, ACT, Australia.}
\author{P. K. Lam}%
\email{ping.lam@anu.edu.au}
\affiliation{Institute of Materials Research and Engineering, Agency for Science, Technology and Research (A*STAR), Singapore, 138634, Singapore.}
 \affiliation{Centre of Excellence for Quantum Computation and Communication Technology, Department of Quantum Science and Technology, Research School of Physics, The Australian National University, Canberra, ACT, Australia.}
\author{S. M. Assad}
\email{cqtsma@gmail.com}
\affiliation{Institute of Materials Research and Engineering, Agency for Science, Technology and Research (A*STAR), Singapore, 138634, Singapore.}
 \affiliation{Centre of Excellence for Quantum Computation and Communication Technology, Department of Quantum Science and Technology, Research School of Physics, The Australian National University, Canberra, ACT, Australia.}




\begin{abstract}
Discrete-variable (DV) quantum key distribution (QKD) based on single-photon detectors and sources have been successfully deployed for long-range secure key distribution. On the other hand, continuous-variable (CV) quantum key distribution (QKD) based on coherent detectors and sources is currently lagging behind in terms of loss and noise tolerance. An important discerning factor between DV-QKD and CV-QKD is the effect of phase noise, which is known to be more relevant in CV-QKD. In this article, we investigate the effect of phase noise on DV-QKD and CV-QKD protocols, including the six-state protocol and squeezed-state protocol, in a thermal-loss channel but with the assumed availability of perfect sources and detectors. We find that in the low phase noise regime but high thermal noise regime, CV-QKD can tolerate more loss compared to DV-QKD. We also compare the secret key rate as an additional metric for the performance of QKD. Requirements for this quantity to be high vastly extend the regions at which CV-QKD performs better than DV-QKD. Our analysis addresses the questions of how phase noise affects DV-QKD and CV-QKD and why the former has historically performed better in a thermal-loss channel.
\end{abstract}

\flushbottom
\maketitle

%
%
\thispagestyle{empty}

\section*{Introduction}

    Quantum key distribution (QKD) enables the sharing of keys between two parties, Alice and Bob. Once a quantum secret key is established, it can later be used by both parties to unlock encrypted communication with total confidentiality. In fact, this form of communication is guaranteed to be secure against an eavesdropper, Eve, by the laws of quantum physics. QKD has become a viable cyber-security technology with increasing interest across government agencies and commercial corporations \cite{nedapredict}. 
    
    The first proposed QKD protocol was based on discrete variables (DV) using two polarization bases, now known as BB84 after its authors Bennett \& Brassard \cite{BENNETT20147}. BB84 and its three-polarization-basis variant, the six-state (6S) protocol, rely on the use of single-photon states and remain robust QKD protocols to this day  \cite{BENNETT20147,simplebb84}. Fifteen years afterward, QKD was extended to continuous variables (CV), initially using entangled multi-photon two-mode squeezed states (TMSV) with low-noise coherent detection  \cite{timsqueeze,ralph2000,hillery2000}. An equivalent scheme known as the squeezed-state protocol was proposed shortly afterwards \cite{assche2001}, requiring preparation only of modulated squeezed states. 
    Subsequently, the GG02 protocol \cite{gross1,gross2,cvhet} with reverse reconciliation and the SRLL02 protocol \cite{silberhorn} based on Gaussian modulation of coherent states eliminated the need for preparing experimentally-challenging squeezed states \cite{Wang:19}. However, the squeezed-state protocol remains relevant due to its ideally better performance and compatibility with certain quantum repeater architectures  \cite{Pirandola20}.  

A comparison between measurement-device-independent (MDI) DV-QKD and CV-QKD protocols, taking into account experimental imperfections was done by Pirandola et. al  \cite{piran2}. This comparison in terms of source and detector technologies was discussed in Ref. \cite{xu, Pirandola2015B}. Subsequently, a technology-independent comparison of DV-QKD and CV-QKD protocols in a noisy channel with ideal sources and detectors have been investigated in Ref. \cite{usenko}. It was shown that CV-QKD protocols are generally robust against noise when loss is low to moderate whereas DV-QKD protocols are superior in very low and strong loss regimes. However, in Ref. \cite{usenko}, the analysis for DV-QKD assumed distinguishability between signal photons and noise photons, leading to favourable results for DV-QKD noise tolerance in the very low loss regime. In practical channels, stray photons from background noise are indistinguishable from the signal photon \cite{erlong}. In addition, the magnitudes of secret key rates of the QKD protocols were ignored as a metric for the comparison. High key rates are an important requirement for a full QKD network to service many users \cite{eleni2016, pan2021}. 

We hypothesize that one of the factors for the consistent historical performance of DV-QKD protocols is mainly due to their robustness to phase noise, which plagues CV-QKD protocols that rely on encoding information in phase as well as amplitude \cite{ralph2000}. We test this hypothesis by introducing a phase noise model consistent with both DV-QKD and CV-QKD. 

In this article, we compare idealized DV-QKD and CV-QKD protocols, the BB84 protocol, the six-state (6S) protocol, and the squeezed-state protocol, by assuming perfect sources, detectors, and reconciliation efficiency in a thermal-loss channel. In doing so, we avoid the dependence on practical implementation and current technological limitations. In the first half of the article, we delve into key-rate comparisons in the thermal-loss channel of QKD protocols. For completeness, we consider the strategy of  ``fighting noise with noise'' for improved performances in both the DV-QKD and the CV-QKD protocols. We also identify gaps, if any, between the ideal performances of these QKD protocols and known bounds on the key capacity in the thermal-loss channel.

In the second half of the article, unlike previous works \cite{piran2, xu, usenko}, we address phase noise in both DV-QKD and CV-QKD, which is a discerning factor for the performance of QKD. We make use of the fact that in the DV-QKD protocol, the thermal-loss and phase noise channels are equivalent to the depolarizing and dephasing channels, respectively. Furthermore, we present results in the combined thermal-loss and phase noise channels. Our work addresses an important question about which QKD protocol performs better by various metrics for a given thermal-loss and phase noise channel. Finally, we discuss and conclude our results in the context of real-world implementations, and possible future directions.

\section{Thermal-loss in QKD}

\begin{figure*}
\centering
 \begin{minipage}[b]{0.49\textwidth}
 \includegraphics{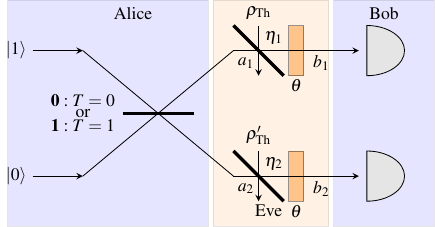}
\caption*{a)}
\end{minipage} \hspace{-0.05\linewidth}
 \begin{minipage}[b]{0.49\textwidth}
\includegraphics{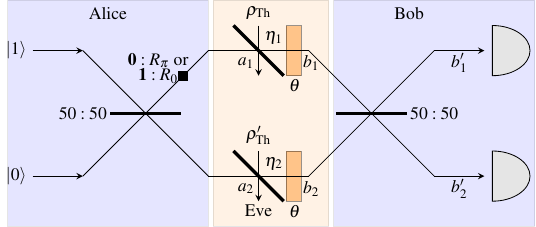}
\caption*{b)}
 \end{minipage} \hspace{0.2\textwidth}
\caption{Dual-rail BB84 protocol in the thermal-loss channel. a) and b) show the rectilinear and diagonal polarization bases as the dual-rail equivalent of the BB84 discrete-variable QKD protocol, respectively. The phase $\theta$ of either bases are affected independently by randomly distributed phase noise $\sigma^2_\theta$. In a), based on which mode (top or bottom) Alice chooses to send a single photon determines the logical bit $\bf{0}$ or $\bf{1}$. In b), based on the phase $0$ or $\pi$ of the rotation $R$ (black square), Alice prepares a logical bit $\bf{0}$ or $\bf{1}$, respectively. }
\label{rect}
\end{figure*}


In this section, we present the security models and secret key rate expressions for the DV-QKD and CV-QKD protocols in the thermal-loss channel. We then present the results of the secret key rate of these calculations.

\subsection{Thermal-loss in the BB84 (and six-state) dual-rail protocol}
\label{bb84security}
We make use of the dual-rail BB84 protocol which is one possible implementation of the original BB84 protocol. In the original BB84 protocol, Alice sends a polarization qubit to Bob with a channel that can support both polarizations. This is equivalent to Alice utilising two quantum channels, each supporting only a single polarization. We present this dual-rail BB84 protocol in Fig. \ref{rect} a) and \ref{rect} b). In the BB84 protocol, Alice prepares a single qubit in either the rectilinear $Z$-basis $\{\ket{0},\ket{1}\}$ or the diagonal $X$-basis $\{\ket{+},\ket{-}\}$. In the rectilinear basis shown in Fig. \ref{rect} a), a logical $\bf{0}$ is prepared by Alice sending a single photon state $\ket{1}$ in the top $a_1$ mode and a vacuum state $\ket{0}$ in the bottom $a_2$ mode. Similarly, a logical $\bf{1}$ is prepared by sending the vacuum state $\ket{0}$ in the top $a_1$ mode and a single-photon state $\ket{1}$ in the bottom $a_2$ mode. The qubits pass through a thermal-loss channel represented by a beamsplitter parameter with transmissivity $0\le \eta_{1,2} \le 1$ and thermal state $\hat{\rho}_{\text{Th}}$ with $N_{\text{Th}}$ thermal photons in the auxiliary port.

Bob, after deciding randomly (discussed in detail later) to measure the $Z$-basis, measures each mode output with single-photon detectors, only accepting single-photon events at $b_{1}$ or $b_{2}$ corresponding to logical $\bf{0}$ or $\bf{1}$. Any other detector events are not counted towards the final key. In the diagonal basis (see Fig. \ref{rect} b)), Alice interferes with a single-photon with the vacuum using a balanced $50:50$ beamsplitter to generate the superposition state $\ket{+}=\frac{1}{\sqrt{2}}(\ket{1}_{a_1}\ket{0}_{a_2}+\ket{0}_{a_1}\ket{1}_{a_2})$ which corresponds to a logical $\bf{1}$ state. A logical $\bf{0}$ corresponds to Alice placing a $\pi$-phase shifter after the beamsplitter and generating the state $\ket{-}=\frac{1}{\sqrt{2}}(\ket{1}_{a_1}\ket{0}_{a_2}-\ket{0}_{a_1}\ket{1}_{a_2})$ to send to Bob. Bob, having randomly decided to measure in the $X$-basis by placing a balanced beamsplitter, measures only single-photon events at $b'_{1}$ or $b'_{2}$ corresponding to logical $\bf{0}$ or $\bf{1}$. We assume the modes pass through the thermal-loss channels with $\eta_1=\eta_2=\eta$ and thermal noise $N_{\text{Th}}$ and no correlations between the two thermal environments. In the final step of the protocol, Bob sends information to Alice about which basis he used. In this reconciliation phase, Alice discards the data that does not match the basis she used to encode her qubits.

The key rate (per channel use) for the BB84 protocol with perfect reconciliation efficiency in the asymptotic limit is \cite{shor2000,renner2008,murta2020}
\begin{equation}
K_{\text{BB84}}=\frac{P_S}{2} (1-h(Q_Z)-h(Q_X)),
\label{key}
\end{equation} where $h(x)=-x \log_2{(x)}-(1-x)\log_2{(1-x)}$ is the binary entropy function, $P_S$ is the success probability of single-photon events, $Q_Z$ and $Q_X$ are the quantum bit error rates (QBERs) of the measurement bases $Z$ and $X$ respectively. Unlike the usual normalization preserving DV channels, the success probability $P_S$ is necessary because the thermal environment adds Gaussian noise, and only single-photon events are counted towards the secret key rate. Here, we assume perfect number-resolving detectors as opposed to click detectors that count all non-vacuum $n>0$ events. 

To calculate $Q_Z$, we consider the probability of a bit-flip if Alice sends a logical $\bf{0}$ (i.e. $\ket{1}_{a_1} \ket{0}_{a_2}$) and Bob detects a logical $\bf{1}$ (i.e. simultaneously detects $\ket{0}_{b_1}$ and $\ket{1}_{b_2}$) with probability given by (see Appendix \ref{appendixa} for full calculations):
\begin{equation}
\begin{split}
    P_{Z,\bf{0}\rightarrow\bf{1}}&=P_{Z,\ket{0}_{a_1} \rightarrow \ket{1}_{b_1}} P_{Z,\ket{1}_{a_2} \rightarrow \ket{0}_{b_2}} \\
    & =\frac{\,N_{\text{Th}}(1+N_{\text{Th}})(1-\eta)^2}{\gamma^{4}},
\end{split}
\end{equation} where $\gamma=1+N_{\text{Th}}-N_{\text{Th}} \eta$.


Bob only accepts the correct bits and the flipped bits using photon-number resolving detectors. Therefore, we normalize by considering the total probability Bob only detects the logical bits in the $Z$-basis. Since we assume the channels are symmetric, $P_{Z,\bf{1}\rightarrow\bf{0}}=P_{Z,\bf{0}\rightarrow\bf{1}}$, the QBER is
\begin{equation}
Q_Z=\frac{P_{Z,\bf{0}\rightarrow\bf{1}}}{P_{Z,\bf{0}\rightarrow\bf{1}}+P_{Z,\bf{0}\rightarrow\bf{0}}}=\frac{P_{Z,\bf{1}\rightarrow\bf{0}}}{P_{Z,\bf{1}\rightarrow\bf{0}}+P_{Z,\bf{1}\rightarrow\bf{1}}},
\label{qz}
\end{equation} where $P_{Z,\bf{0}\rightarrow\bf{0}}=P_{Z,\ket{1}_{a_1} \rightarrow \ket{1}_{b_1}} P_{Z,\ket{0}_{a_2} \rightarrow \ket{0}_{b_2}}$ and $P_{Z,\bf{1}\rightarrow\bf{1}}=P_{Z,\ket{0}_{a_1} \rightarrow \ket{0}_{b_1}} P_{Z,\ket{1}_{a_2} \rightarrow \ket{1}_{b_2}}$ are the probabilities of Bob detecting the same bits that Alice sent after passing through the channel. The probability of an event (or success) is given by:
\begin{equation}
\begin{split}
P_S&=P_{Z,\bf{0}\rightarrow\bf{1}}+P_{Z,\bf{0}\rightarrow\bf{0}}\\
&=\frac{\eta + 2\,N_{\text{Th}}(1+N_{\text{Th}})(1-\eta)^2}{\gamma^{4}}.
\end{split}
\label{success}
\end{equation}

To calculate $Q_X$, we consider the bit-flips in the $X$ basis. In this case, the modes $a_1$ and $a_2$ are entangled because of the balanced beamsplitter (see Fig. \ref{rect} b)). Similar to above, we obtain the QBER, for the $X$ bases as
\begin{equation}
Q_X=\frac{P_{X,\bf{0}\rightarrow\bf{1}}}{P_{X,\bf{0}\rightarrow\bf{1}}+P_{X,\bf{0}\rightarrow\bf{0}}}.
\label{qx1}
\end{equation}
We find due to symmetry that the probabilities for the diagonal basis are the same as for the rectilinear basis and it follows that $Q_X=Q_Z$, simplifying the key rate equation. We make use of Eqs. (\ref{key}), (\ref{qx1}), and (\ref{success}) to calculate the key rate in the asymptotic limit. 

    Conditioned on the outcome with probability $P_S$, it can be shown that the density matrix after the thermal-loss channel is, in fact, a depolarized state (see Appendix \ref{depolar}):
\begin{equation}
\begin{split}
\tilde{\rho}^B := \frac{\rho^B}{P_S} &= \frac{\eta}{\eta + 2\,N_{\text{Th}}(1+N_{\text{Th}})(1-\eta)^2} \rho^A \\
&+ \frac{N_{\text{Th}}(1+N_{\text{Th}})(1-\eta)^2}{\eta + 2\,N_{\text{Th}}(1+N_{\text{Th}})(1-\eta)^2}\I,
\end{split}
\end{equation} where $\rho^A$ is Alice's initial density matrix.
This represents a depolarizing channel \cite{nielsen}
\begin{equation}\label{eq:depolarizing-channel1}
  \rho \to (1-\lambda)\rho + \frac{\lambda}{2}\I
\end{equation}
with depolarizing parameter
\begin{equation}\label{eq:depolarising-parameter1}
\lambda = \frac{2\,N_{\text{Th}}(1+N_{\text{Th}})(1-\eta)^2}{\eta + 2\,N_{\text{Th}}(1+N_{\text{Th}})(1-\eta)^2},
\end{equation}
which tends to 1 as $\eta \to 0$ or $N_{\text{Th}} \to \infty$, as expected. 

A property of the depolarizing channel is that the error rate is the same in all bases:
\begin{equation}\label{eq:thermal-qber}
Q_{X,Y,Z} = \frac{\lambda}{2} = \frac{N_{\text{Th}}(1+N_{\text{Th}})(1-\eta)^2}{\eta + 2N_{\text{Th}}(1+N_{\text{Th}})(1-\eta)^2},
\end{equation}
which can be seen from Eq. \eqref{eq:depolarizing-channel1}. 
In establishing this equivalence between the thermal-loss and depolarizing channel, we extend our analysis to the six-state protocol which makes use of an additional basis $Y$ with QBER $Q_Y$. The key rate for the 6S protocol is given by \cite{murta2020}:
\begin{equation}
\label{sixstate}
    K_{\text{6S}}=\frac{P_S}{2}(1-H(\Lambda_{00})-H(\Lambda_{01})-H(\Lambda_{10})-H(\Lambda_{11})),
\end{equation} where $H(x)=-x \log_2{x}$, and
\begin{equation}
    \begin{split}
        \Lambda_{00}&=1-\frac{Q_X+Q_Y+Q_Z}{2}, \\
        \Lambda_{01}&=\frac{Q_X+Q_Y-Q_Z}{2},\\
        \Lambda_{10}&=\frac{-Q_X+Q_Y+Q_Z}{2},\\
        \Lambda_{11}&=\frac{Q_X-Q_Y+Q_Z}{2},
    \end{split}
\end{equation} where the factor of $1/2$ is to normalize the key-rate to per channel use. In the thermal-loss channel, the QBER $Q_X=Q_Y=Q_Z$ is symmetric. However, as we will see when phase noise is introduced, the QBER of the three bases can be asymmetric.

\subsection*{Lower bounds of BB84 and 6S protocols in the thermal-loss channel}
\label{lowerbounds}
Introducing random bit flips at Alice before the error processing increases the performance of BB84 in a noisy channel and sets a tighter lower bound on the key rate \cite{BB84noise}. In this extension of the BB84 protocol which we denote as NBB84, the key rate equation depends on Alice's added bit-flip probability $q$ (or trusted bit-flips). Following Ref. \cite{BB84noise}, we make use of the QBER for the thermal-loss channel in Eq. \eqref{eq:thermal-qber} and maximize the key rate with respect to $q$. We note that the six-state protocol (with and without trusted bit-flips) can tolerate higher QBER than the BB84 protocol \cite{BB84noise}. 
Similarly, the lower bound on the secret key rate of the 6S protocol is likewise calculated by introducing bit-flips at Alice which increases the QBER tolerance of the channel \cite{BB84noise}.
\label{lowerbounds}

\begin{figure*}
\centering
 \begin{minipage}[b]{0.49\textwidth}
\centering
\includegraphics{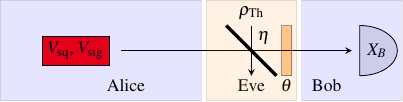}
            \caption*{a)}
 \end{minipage} \hfill
  \begin{minipage}[b]{0.49\textwidth}
   \centering
\includegraphics{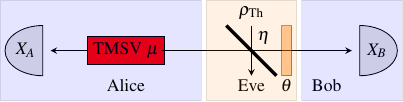}
         \caption*{b)}
 \end{minipage}
\caption{Squeezed-state protocol with homodyne detection in the thermal-loss channel. The phase shifter $\theta$ represents the phase noise $\sigma_\theta^2$. Shown in a) is the equivalent prepare and measure squeezed-state protocol and in b) is the entanglement-based version of the squeezed-state protocol. }
\label{Sqzhom}
\end{figure*}
\subsection{Thermal-loss in the squeezed state protocol}
\label{cvqkdsecurity}

In the squeezed-state protocol in a prepare-and-measure (PM) scheme presented in Fig. \ref{Sqzhom} a), Alice introduces a modulation signal in either the $\hat{X}=\hat{a}+\hat{a}^\dagger$ or $\hat{P}=-i(\hat{a}-\hat{a}^\dagger)$ quadrature (randomly chosen) a squeezed state with $V_{\text{sq}}$ with Gaussian distribution centered at $0$ with variance $V_{\text{sig}}$. In the equivalent entanglement-based (EB) scheme presented in Fig. \ref{Sqzhom} b), Alice performs a homodyne measurement on one mode of a shared two-mode squeezed vacuum state (TMSV) where the other mode passes through the channel $\mathcal{E}$, and Bob performs a homodyne measurement \cite{raulsanchez}. 
The parameter transformation between the PM and EB schemes is:
\begin{equation}
\begin{split}
V_{\text{sq}}&=1/\mu, \\
V_{\text{sig}}&=\frac{\mu^2-1}{\mu},
\end{split}
\end{equation} where $\mu=V_{\text{sq}}+V_{\text{sig}}$ is the quadrature variance of $\hat{X}$ and $\hat{P}$ of the TMSV source in EB scheme. The following key rate calculations are in the EB scheme. In the asymptotic regime of infinite keys, Eve's most powerful attack is a collective attack. Security proofs in this regime for this protocol are based on reduction of coherent attacks to collective attacks for infinite dimensions and on the optimality of Gaussian attacks \cite{wolf,patron,nava}. The secret key rate against collective attacks in the asymptotic regime with reverse reconciliation is given by \cite{laudenbach}
\begin{equation}
K_{\text{RR}}=\beta I_{\text{AB}}-\chi_{\text{EB}},
\label{rrkeyrate}
\end{equation} where $\beta$ is the reconciliation efficiency, $I_{\text{AB}}$ is the mutual information between Alice and Bob, and $\chi_{\text{EB}}$ is the Holevo information between Bob and Eve. In a Gaussian channel, the quadrature covariance matrix between Alice and Bob is \cite{raulsanchez}:
\begin{equation}
\begin{split}
&\gamma_{\rm AB}=
\begin{pmatrix}
a\mathbb{I} & c\sigma_{z} \\ c\sigma_{z} & b\mathbb{I}
\end{pmatrix}
=
\begin{pmatrix}
\gamma_\text{A} & \sigma_{\text{AB}} \\ \sigma_{\text{AB}} & \gamma_\text{B}
\end{pmatrix} = \\
&
\begin{pmatrix}
(V_\text{A}+1)\mathbb{I} & \sqrt{\eta(V_\text{A}^2+2V_\text{A})}\sigma_{z} \\ \sqrt{\eta(V_\text{A}^2+2V_\text{A})}\sigma_{z} & V_B\mathbb{I}
\end{pmatrix},
\end{split}
\label{covariance}
\end{equation}
where $V_\text{A}=\mu-1$ and $V_B=\eta(V_\text{A}+1+\chi)$ are the TMSV variances measured by Alice and Bob (respectively), $\chi=\frac{(1-\eta)(2 N_\text{Th}+1)}{\eta}$ is the noise of the thermal-loss channel, $\mathbb{I}=\text{diag}(1,1)$ is the unity matrix and $\sigma_{z}~=~\text{diag}(1,-1)$ is the Pauli-Z matrix. 
We choose homodyne detection (also known as ``switching") at Bob, in which Bob switches between $\hat{X}$ or $\hat{P}$ quadrature measurements. 

In the squeezed state protocol with homodyne detection, the mutual information is given by:
\begin{equation}
I_{\text{AB}}^{\text{hom}}=\frac{1}{2}  \log_2{\Big( \frac{V_\text{B}}{V_{\text{B|A}}} \Big)},
\end{equation} where $V_{\text{B|A}}=b-\frac{c^2}{a}$. 
The Holevo information between Bob and Eve for the collective attack is given by 
\begin{equation}
\chi_{\text{EB}}=S(\text{E})-S(\text{E|B}),
\label{eve}
\end{equation} where $S(\text{E})$ is Eve's information and $S(\text{E|B})$ is Eve's information conditioned on Bob's measurement. In Eve's collective attack, Eve holds a purification of the state between Alice and Bob with entropy given by
\begin{equation}
S(\text{E})=S(\text{AB})=G[(\lambda_1-1)/2]+G[(\lambda_2-1)/2],
\label{pure}
\end{equation} where $G(x)=(x+1) \log_2{(x+1)}-x \log_2{x}$ and $\lambda_{1,2}$ are the symplectic eigenvalues of the covariance matrix $\gamma_{\text{AB}}$ given by
$\lambda_{1,2}^2=\frac{1}{2} [\Delta \pm \sqrt{\Delta^2-4\mathcal{D}^2}]$, where 
$\Delta=\text{Det}(\gamma_\text{A})+\text{Det}(\gamma_\text{B})+2\text{Det}(\sigma_{\text{AB}})$, and
$\mathcal{D}=\text{Det}(\gamma_{\text{AB}})$.
The conditional covariance matrix of Alice's mode after the homodyne detection by Bob is
\begin{equation}
\Gamma_{\text{A}|b}=
\begin{pmatrix}
\mu-\frac{\eta(\mu^2-1)}{\eta\mu+(1-\eta)(2N_{\text{Th}}+1)} & 0  \\ 
0 & \mu
\end{pmatrix}.
\label{consq}
\end{equation}
Therefore, Eve's entropy conditioned on Bob's measurement $S(\text{E|B})=S(\text{A}|b)$ is given by $G[(\lambda_3-1)/2]$ where $\lambda_3$ is the symplectic eigenvalue of $\Gamma_{A|b}$. 

Introducing trusted noise before Bob's homodyne measurement can help extend a high-noise thermal-loss channel. In this extension of the squeezed-state protocol which we denote NSqz-Hom, trusted Gaussian noise $\xi_{\text{B}}$ is added before post-processing on Bob's homodyne measurement data  \cite{raulsanchez}. The effect is that Eve's information decreases more than the mutual information between Alice and Bob (see Appendix \ref{appendixwithnoise} for calculations), thus increasing the secret key rate of the protocol. Similarly, heterodyne detection at Bob has the same effect of introducing additional noise, thereby extending secure communication distance in a thermal-loss channel  \cite{garcianoise}. 

{
\section{Phase noise in QKD}

\newcommand{\sigmatilde}{\tilde{\sigma}}
We consider a standard model of bosonic phase noise, known also as dephasing, phase-diffusion, or phase-damping. This channel represented by $\theta$ on the right of Fig. \ref{rect} applies rotation by a random angle $\theta$ to the bosonic state according to a classical distribution $f(\theta)$, giving the transformation
\begin{equation}
    \hat{\rho} \longmapsto \int_{-\pi}^{\pi}d\theta f(\theta) e^{i\ah\hc\ah\theta}\hat{\rho} e^{-i\ah\hc\ah\theta}.
\end{equation}

Since $\ah\hc\ah$ is the number operator, a given rotation $\theta$ applies a phase $e^{in\theta}$ to each Fock state $\ket{n}$, equivalently described by the transformation
\begin{equation}\label{eq:phase-transform}
  \ah\hc \longmapsto e^{i\theta} \ah\hc.
\end{equation}
The canonical phase distribution is the \emph{wrapped normal distribution}, which models the random diffusion of an angle and accurately represents the physical process of phase diffusion \cite{nielsen}. Birefringence may produce this behaviour in polarisation-based implementations of the six-state and BB84 protocols or in time-bin implementations, phase drift in between the interferometers at either end \cite{dvphasenoise}.

The phase shift $\theta$ (assumed here to have mean zero) is normally distributed over the whole real line:

\begin{equation}
    \tilde{f}_{WN}(\theta) = \frac{1}{\sigma_\theta\sqrt{2\pi}} e^{-\theta^2/2\sigma^2_\theta}:\qquad\theta\in\mathbb{R},
\end{equation}
which we can `wrap' into a single $2\pi$ interval by summing the contributions from equivalent angles:
\begin{equation}
    f_{WN}(\theta) = \frac{1}{\sigma_\theta\sqrt{2\pi}} \sum_{k=-\infty}^{\infty}e^{-(\theta + 2\pi k)^2/2\sigma^2_\theta}:\theta\in[-\pi,\pi].
\end{equation}
The variance $\sigma^2_\theta$ of $\theta$ over the whole real line is in general \emph{not} its variance when wrapped; however, the two distributions approach each other in the limit of small variance.

\newcommand{\eit}{e^{i\theta}}
\newcommand{\btheta}{\bm{\theta}}
\newcommand{\urot}{\hat{U}_{\btheta}}
\newcommand{\bk}{\mbf{k}}

The corresponding qubit transformation of the phase noise ignoring the thermal-loss ($\eta=1$) is $\rho_{jk} \mapsto e^{i\theta_j}e^{-i\theta_k}\rho_{jk}$, which may be expressed as $\hat{\rho} \rightarrow \urot\hat{\rho}\urot\hc$ where $\urot = \diag(e^{i\theta_i}, e^{i\theta_j})$. If $\btheta$ is drawn from a distribution $f(\btheta)$, the qubit channel becomes
\begin{equation}
    \hat{\rho} \to \mean{\urot\hat{\rho}\urot\hc} = \int_{\btheta} f(\btheta) \urot\hat{\rho}\urot\hc\;d\btheta
    \label{rhophasenoise}
\end{equation}
where $\mean{\cdot}$ denotes expected value. If $\{\theta_j\}$ are independent, the corresponding transformation of off-diagonal terms ($i\neq j$) is
\begin{equation}
    \rho_{jk} \longmapsto \mean{e^{i\theta_j}e^{-i\theta_k}} \rho_{jk} = \mean{e^{i\theta_j}}\mean{e^{-i\theta_k}} \rho_{jk} = \Bar{r}_j\bar{r}_k^*\rho_{jk}
\end{equation}
where $\Bar{r}_j := \mean{e^{i\theta_j}}$ is the so-called `circular mean' of $\theta_j$, given for the wrapped normal distribution:

  \begin{equation}
        \label{wrapped}
    \Bar{r}_j = e^{-\sigma^2_\theta/2}.
  \end{equation}

Diagonal entries remain unchanged:
\[\rho_{jj} \longmapsto \mean{e^{i\theta_j}e^{-i\theta_j}}\rho_{jj} = \rho_{jj}.\]
 If $\{\theta_j\}$ are identically distributed then all have the same (real) circular mean $\Bar{r}$ and we obtain a (generalised) dephasing channel
\begin{equation}\label{eq:hdqkd-dephasing}
  \hat{\rho} \longmapsto \hat{\rho}_{\text{dephased}}=\bmtx{\rho_{00} & \Bar{r}^2\rho_{01} \\ \Bar{r}^2\rho_{10} & \rho_{11}}
\end{equation}
which always sends single-photon inputs to single-photon outputs, unlike the thermal-loss channel. By leaving diagonal entries unchanged, Eq. \eqref{eq:hdqkd-dephasing} introduces \emph{no error} in the Z basis. Common DV-QKD protocols such as the (generalised) BB84 and six-state protocols make use of additional bases ($X$ and $Y$) which are \emph{unbiased} with respect to the Z basis.

The extension to the combined thermal-loss phase-noise rail presented in Fig. \ref{rect} can be obtained by composing the separate depolarization and dephasing channels described in Eqs. \eqref{eq:depolarizing-channel1} and \eqref{eq:hdqkd-dephasing}, giving
\[\hat{\rho} \longrightarrow (1-\lambda)\bmtx{\rho_{00} & \Bar{r}^2\rho_{01} &  \\ \Bar{r}^2\rho_{10} & \rho_{11} }  + \frac{\lambda}{2}\I.\]
The corresponding error rates are thus $(1-\lambda)\hat{\rho}_{\mathrm{dephased}} + \frac{\lambda}{2} \I$, i.e.
\begin{equation}\label{eq:combined-qber}
  \begin{aligned}
    Q_Z &= \left(\frac{\lambda}{2}\right), \\[2ex]
    Q_{X,Y} &= \left(\frac{1}{2}\right)\left[(1-\lambda)(1-\Bar{r}^2) + \lambda\right],
  \end{aligned}
\end{equation}
with $\Bar{r}^2$ and $\lambda$ given by Eqs. \eqref{wrapped} and \eqref{eq:depolarising-parameter1} respectively. The probability of success $P_S$ remains the same as for the thermal-loss channel in Eq. \eqref{success}, as the subsequent dephasing does not affect which states are discarded. The key rate for the BB84 and 6S protocols are straightforward to calculate from these QBERs and using Eqs. \eqref{key} and \eqref{sixstate}, respectively. 
\begin{figure*}[t!]
\centering
\includegraphics[scale=0.61]{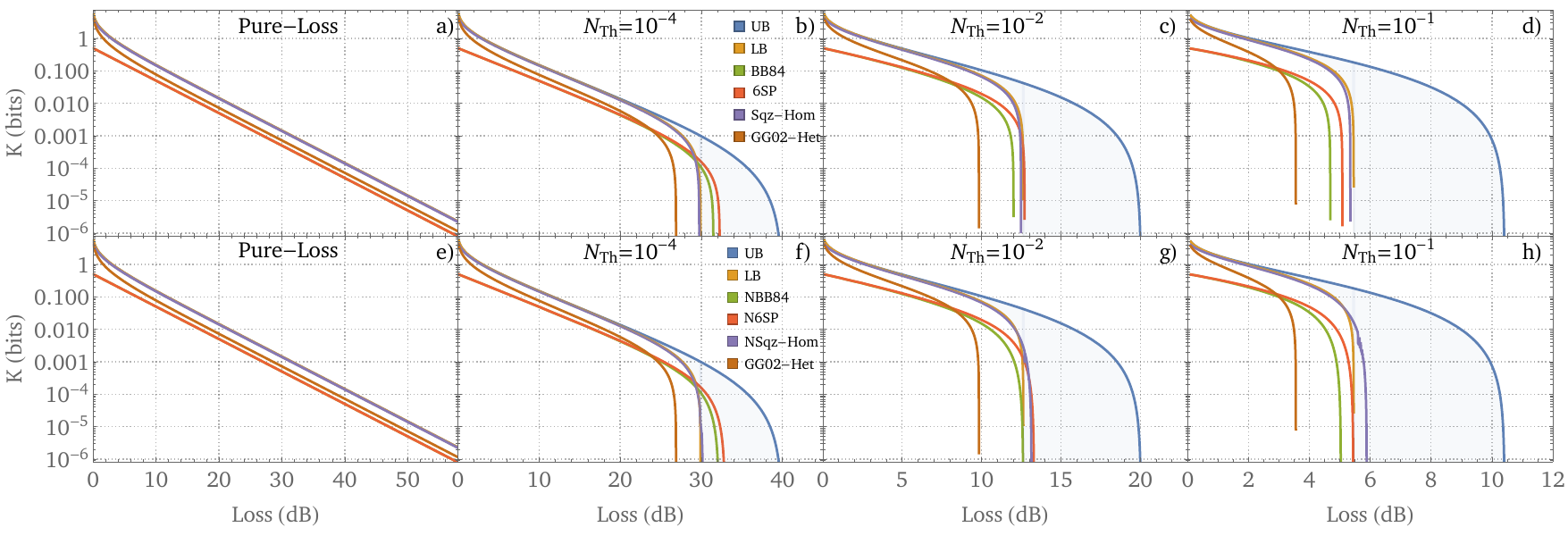}
\caption[l]{Secret key rate per polarization channel in a thermal-loss channel for increasing noise $N_{\text{Th}}$. Figures a)-d) and e)-h) show the QKD versions of the protocols without and with trusted noise, respectively. For comparison, we also include the GG02 in all figures. For the pure-loss channel, the Sqz-Hom and NSqz-Hom essentially overlap with the PLOB bound for the chosen squeezing of $15~\text{dB}$. Next in b) and f) with some noise in the thermal-loss channel means that the BB84, NBB84, 6S and N6S protocols outperform the CV-QKD protocols. As shown in c), d), g) and h) as more thermal noise is present, the Sqz-Hom and NSqz-Hom outperform BB84, NBB84, 6S, and N6S. In particular, Sqz-Hom saturates the lower bound (LB). Lastly, NSqz-Hom is by far the best protocol in a high noise regime as shown in h) but far from the upper bound (UB). }
\label{lowthermal}
\end{figure*}

Turning to CV-QKD, the phase noise channel is given by the same random rotation $\hat{a}\to\hat{U} \hat{a} \hat{U}^\dagger = e^{-i\theta} \hat{a}$, with $\theta$ drawn from a distribution as in Eq. (\ref{rhophasenoise}). This models statistical error in the phase, regardless of the chosen CV-QKD implementation scheme. For a given $\theta$, a coherent state $\ket{\alpha}$ becomes $\ket{\alpha e^{-i\theta}}$. 

For the squeezed-state protocol, the combined thermal-loss and phase noise channel leads to the following covariance matrix:
\begin{equation}
\begin{split}
&\gamma_{\rm AB}=
\begin{pmatrix}
a\mathbb{I} & c\sigma_{z} \\ c\sigma_{z} & b\mathbb{I}
\end{pmatrix}
 \\
&
=\begin{pmatrix}
\mu\mathbb{I} & \sqrt{\eta_\text{I}(\mu^2-1})\sigma_{z} \\ \sqrt{\eta_\text{I}(\mu^2-1})\sigma_{z} & (\eta_\text{I} (\mu+ \chi_\text{I})\mathbb{I}
\end{pmatrix} \\
&
=\begin{pmatrix}
\mu\mathbb{I} & \Bar{r}\sqrt{\eta (\mu^2-1})\sigma_{z} \\ \Bar{r}\sqrt{\eta(\mu^2-1})\sigma_{z} & (\eta \mu+(1-\eta) (2 N_\text{Th}+1))\mathbb{I}
\end{pmatrix},
\end{split}
\label{covariance2}
\end{equation} where the inferred transmittance $\eta_I=\eta \Bar{r}^2=\eta e^{-\sigma^2_\theta}$ and noise $\chi_I=\frac{(1-e^{-\sigma^2_\theta})\eta\mu+(1-\eta)(2 N_\text{Th}+1)}{\eta e^{-\sigma^2_\theta}}$ are derived in Appendix \ref{bestphasenoise}. Since phase noise wraps squeezed states around the axis, the average state of modulated squeezed states remains a thermal state but the correlations $\braket{X_A X_B}$ are now reduced by the factor $\Bar{r}=e^{-\sigma^2_\theta/2}$. It is straightforward to calculate the SKR using Eq. (\ref{covariance2}) and the equations in the previous section.

We note that in the regime where $\sigma^2_\theta$ is large, the phase diffusion channel becomes non-Gaussian \cite{wilde-bosonic}. Since we are considering the squeezed-state protocol with coherent detection, we make use of Eq. \eqref{eve} to calculate a lower bound on the key rate. It is left for future work to determine the optimal protocol in the non-Gaussian phase diffusion channel}.

\section{Comparison of QKD protocols}

\subsection{Without phase noise $\sigma_\theta^2=0$}
\label{compare}
In CV-QKD, information is encoded in the $\hat{X}$ and/or $\hat{P}$ quadratures in one polarization with access to an infinite Hilbert space. Conversely, in DV-QKD, information is encoded in one or more polarization basis in a $2$-dimensional Hilbert space. To make a fair comparison, we assume that Alice uses one polarization basis asymptotically close to $100\%$ of the time (the "computational" basis). The other basis is only measured to characterize channel parameters and the QBER. We make a similar assumption for the squeezed state protocols in the sense that Bob rarely switches the quadrature he measures to characterize the anti-squeezing and determine whether Eve tampered with the shared EPR state, thus removing the usual sifting factor of $1/2$ that comes with switching. 

We also make the following ideal assumptions about the CV-QKD and DV-QKD protocols in the thermal-loss channel: (i) single-photon and laser sources are perfect (ii) detectors that are used are ideal with detector efficiencies $\eta_\text{d}=1$ and detector noise $\xi_{\text{det}}=0$ (except for intentionally adding {\it trusted} noise in the ``fighting noise with noise" protocol) (iii) all channel parameters have been estimated with no statistical error (iv) all channel noise is attributed to Eve (v) reverse reconciliation efficiency is perfect with $\beta=1$ and error correction efficiency is perfect for both CV-QKD and DV-QKD (vi) all security analysis is in the asymptotic limit. Our simplified analysis here is valid in the ideal situation where squeezed and coherent states are only affected by loss, thermal noise and (in the next section) phase noise. 

A fundamental benchmark for QKD is the Pirandola, Laurenza, Ottaviani and Banchi (PLOB) bound of a quantum channel $\mathcal{E}$. The PLOB bound establishes upper bounds on two-way secret key capacities $C(\mathcal{E})$ and on point-to-point QKD protocols \cite{plob}. It is known that the PLOB bound in a {\it pure-loss} channel is tight and the squeezed-state protocol, under some conditions, saturates this bound in the limit of infinite squeezing \cite{Pirandola20}. However, there is a gap between the best lower bound and the best upper bound of the secret key capacity of channels with thermal noise. Pirandola et. al determined lower bounds (LB) and upper bounds (UB) on the secret key capacity $C(\eta, N_{\text{Th}})$ of the thermal loss channel where $N_\text{Th}$ is the thermal noise and $\eta$ is the transmissivity of the thermal-loss channel \cite{plob,rci}. The lower bound is given by the reverse coherent information of the thermal-loss channel and the upper bound by the Gaussian relative entropy of entanglement (of the Choi state in the thermal-loss channel):
\begin{equation}
\begin{split}
    & C(\eta,N_{\text{Th}}) \ge -\log_2{[1-\eta]}-G(N_{\text{Th}})=K_{\text{Lower}}, \\ 
    & C(\eta,N_{\text{Th}}) \le -\log_2{[(1-\eta)\eta^{N_{\text{Th}}}]}-G(N_{\text{Th}})\\
    &\qquad \qquad \qquad \qquad \qquad \qquad \qquad \qquad=K_{\text{Upper}},
\end{split}
\end{equation} given for non-entanglement breaking channels $N_\text{Th} < \eta/(1-\eta)$ and where $G(x)=(x+1) \log_2{(x+1)}-x \log_2{x}$.
\begin{figure*}[t!]
\centering
\includegraphics[scale=0.41]{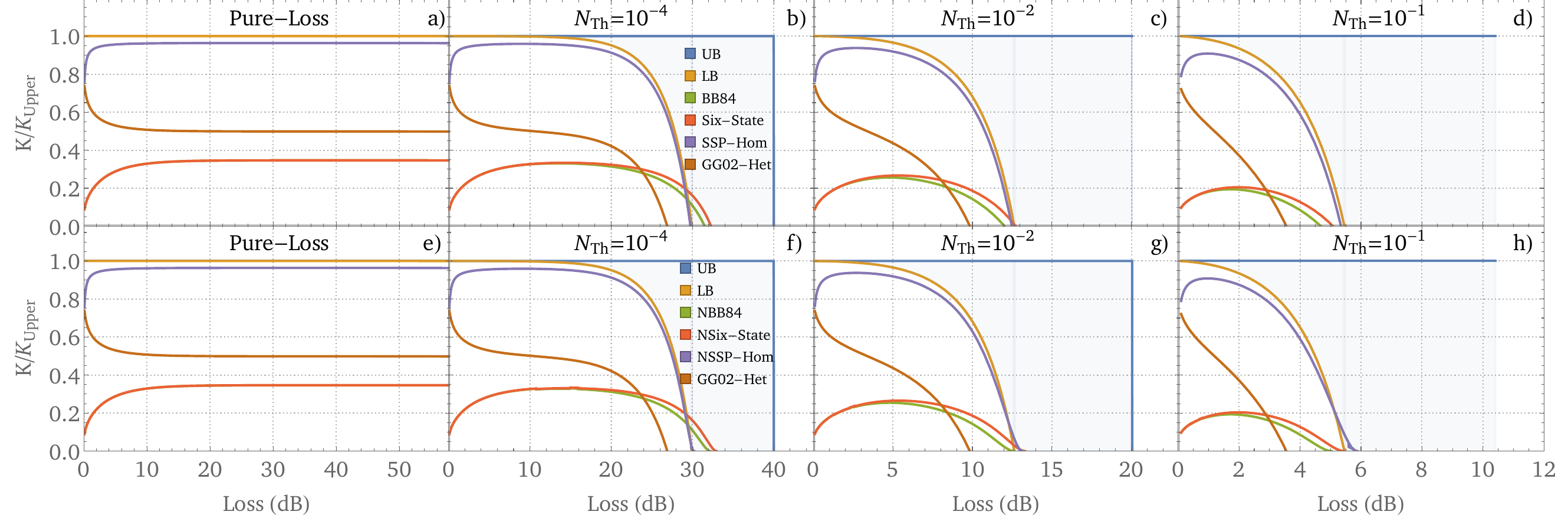}
\caption{Same as Fig. \ref{lowthermal} but to benchmark the performance of the different protocols, we normalise the key rates by the upper bound $K_{\text{Upper}}$. We note that the Sqz-Hom protocol is very close to the LB bound. For clarity in the pure-loss channel, from top to bottom is UB=LB, SSP-Hom, GG02-Het, Six-State=BB84. }
\label{normalized}
\end{figure*}

\begin{figure*}[t!]
\centering
\includegraphics[scale=0.4]{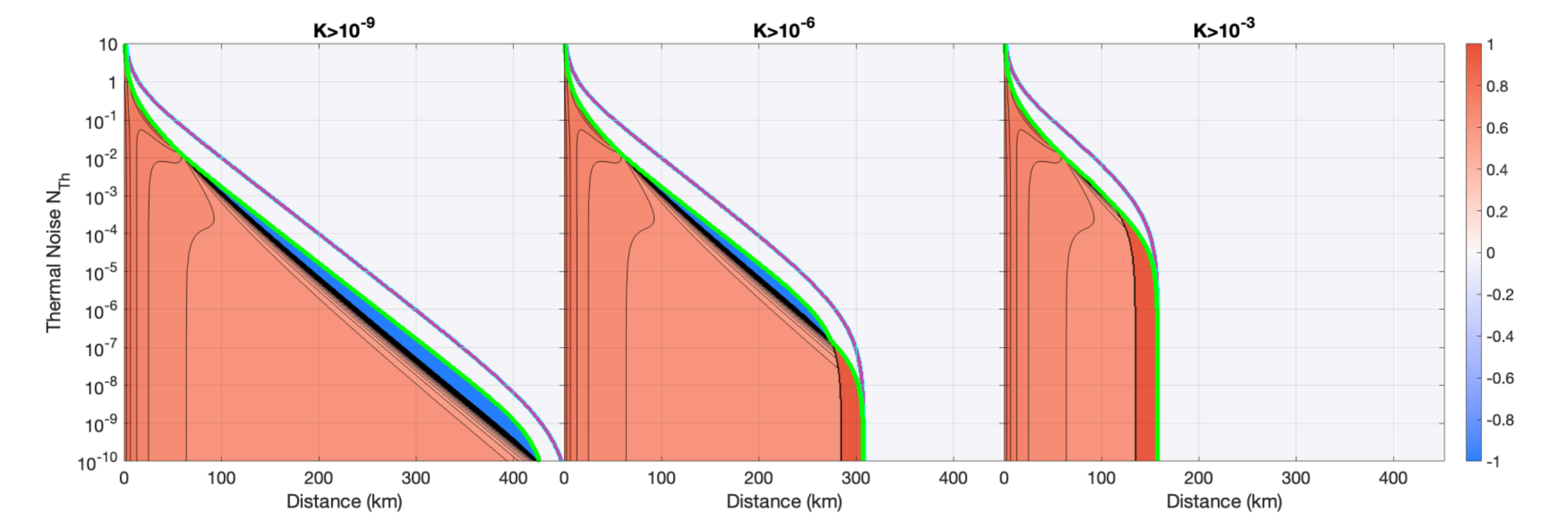}
\caption{Comparison of $\tilde{K}^{\text{CV:DV}}$ for protocols for a set of thermal-loss channel parameters. Blue regions indicate where the 6S protocol has a higher key rate than Sqz-Hom and conversely, red regions are where the Sqz-Hom has higher key rates than the 6S protocol. Given a minimum key rate requirement, we compare the protocols which operate the best for single-channel use QKD. The 6S protocol covers a small region of intermediate noise and loss as seen in the first two subfigures. The green line indicates where the QKD protocols can operate up to the minimum key rate. The rest of the parameter space is covered by the squeezed state protocol. For high key rates in the rightmost subfigure, the 6S protocol always performs worse than Sqz-Hom. The purple line is the upper bound (UB) of the key capacity in the thermal-loss channel. The red region in the middle and right subfigures are regions where only the Sqz-Hom can achieve the minimum key rate.}
\label{optimalchannel}
\end{figure*}

\begin{figure}[t!]
\centering
\includegraphics[width=0.5\textwidth,trim=10mm 50mm 10mm 50mm, clip=true]{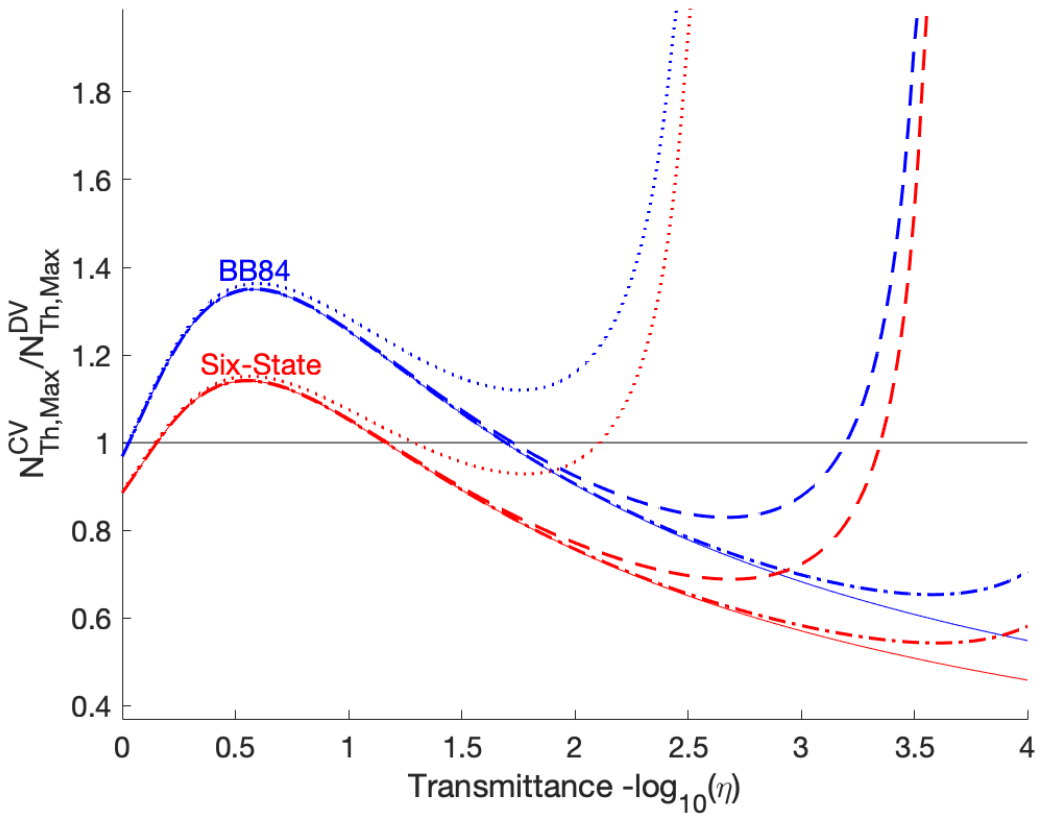}
\caption{Ratio of maximum noise tolerance of the Sqz-Hom protocol with BB84 (blue) and Six-State (red) protocols. Shown are numerical results for $K_0$ of $10^{-3}$ (dotted line), $10^{-4}$ (dashed line), $10^{-5}$ (dotted-dashed line) and $10^{-10}$ (solid line). The protocol parameters of CV-QKD and DV-QKD are the same as those in Fig. \ref{lowthermal}.}
\label{ratioofnoise}
\end{figure}
We present our results in Fig.  \ref{lowthermal} a)-d) for the secret key rate per polarization channel based on calculations of the BB84 protocol, the 6S protocol, the GG02 protocol (see Appendix \ref{appendixgg02}  calculations), and the squeezed state with homodyne (Sqz-Hom) protocols in the thermal-loss channel for various thermal noise parameters. We note that since we make use of the dual-rail BB84 protocol which is one possible implementation of the BB84 protocol, the key rate equation for the DV-QKD protocols has been divided by $2$ into units of symbols per polarization channel. For the Sqz-Hom protocol, we choose a practically achievable squeezing $V_{\text{sq}}$ of $15$~dB \cite{squeezed15}. We note that adding more squeezing only adds a very small improvement to the key rates (see Appendix \ref{bb84vssqueeze} for more details). In the limit of infinite squeezing, the secret key rate of the Sqz-Hom would approach the lower bound (LB) of the secret key capacity in the thermal-loss channel as shown most clearly in Fig. \ref{lowthermal} b) and in a pure-loss channel as shown in Fig. \ref{lowthermal} a). The BB84 and 6S protocols surpass the lower bound in an intermediate thermal-noise regime as shown in Fig. \ref{lowthermal} b). In Fig. \ref{lowthermal} e)-h), we present the ``fighting noise with noise" versions of the protocols. For the squeezed state protocol with homodyne detection  (NSqz-Hom) 
with $15$~dB squeezing we optimized with respect to the trusted noise $\xi_{B}$. As shown in Fig. \ref{lowthermal} g) and h) surpasses the LB for high thermal noise. In addition, the secret key rate of the ``fighting noise with noise" versions of the DV-QKD protocols, the NBB84 and N6S protocols are optimized with respect to the added bit-flips by Alice $q$ and a slight advantage is obtained as shown in Fig. \ref{lowthermal} c). In Fig. \ref{normalized}, to benchmark the performance of the different protocols, we normalized the key rate to the upper bound of the secret key capacity. 

\begin{figure*}[t!]
\centering
        \vskip\baselineskip
        \begin{subfigure}[b]{1\textwidth}   
            \centering 
            \includegraphics[width=1.17\textwidth,trim=45mm 0mm 40mm 0mm, clip=true]{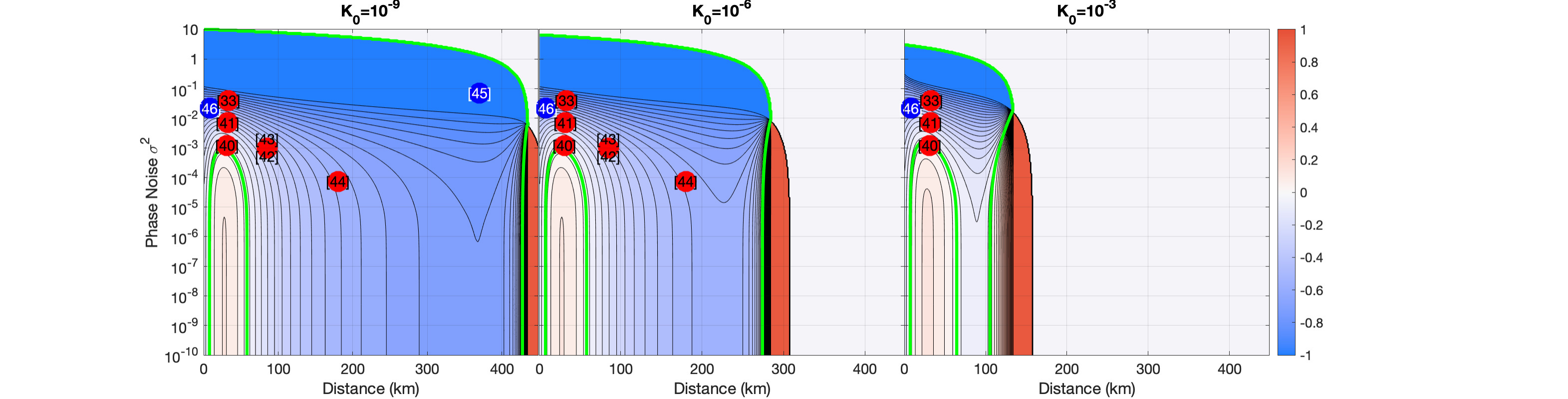}
            \caption{{\small  Contour plot of $\tilde{N}_{\text{Th}}^{\text{CV:DV}}$ as a function of the phase noise and distance (or loss) for the Sqz-Hom and 6S protocol. The green line indicates the point at which both protocols tolerate the same amount of thermal noise. In the white regions to the right-hand side, neither one of the protocols tolerate any thermal noise. In the red region, only the Sqz-Hom protocol tolerates thermal noise. We also show current state-of-the-art CV-QKD (red circles) and DV-QKD (blue circles) protocols. }}
            \label{fig:mean and std of net34}
        \end{subfigure}
         \vskip\baselineskip
        \begin{subfigure}[b]{1\textwidth}   
            \centering 
            \includegraphics[width=1.17\textwidth,trim=45mm 0mm 40mm 0mm, clip=true]{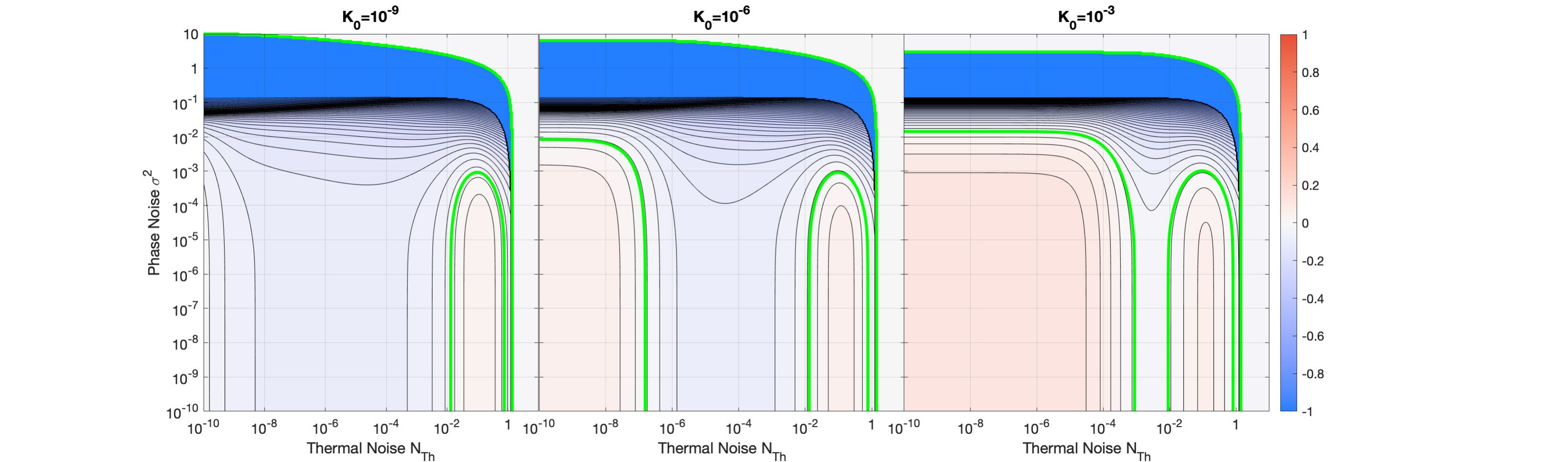}
            \caption{{\small  Contour plot of $\tilde{L}^{\text{CV:DV}}$ or $\tilde{D}^{\text{CV:DV}}$ as a function of the phase noise and thermal noise for the Sqz-Hom and 6S protocol. The green line indicates the point at which both protocols tolerate the same amount of loss. In the white regions to the right-hand side, neither one of the protocols tolerate any loss. }}     
            \label{fig:mean and std of net44}
        \end{subfigure}
        \caption{}
        \label{fig:mean and std of nets}
\label{phasenoise}
\end{figure*}

We compare the protocols by plotting the parameter:
\begin{equation}
    \tilde{K}^{\text{CV:DV}}=\frac{K_{\text{Sqz-Hom}}-K_{\text{6S}}}{\text{Max}[K_{\text{Sqz-Hom}},K_{\text{6S}}]},
\end{equation} for channel parameters of standard optical fibre of loss $0.2~\text{dB/km}$ with distance $D=-50 \log_{10}{(\eta)}~\text{ km}$ and $N_{\text{Th}}$ in Fig. \ref{optimalchannel}. The key rate $(K_{\text{Sqz-Hom}},K_{\text{6S}})>K_0$ where $K_0$ is the minimum required key rate. When the squeezed-state protocol is significantly higher in key rates $\tilde{K}=1$, and conversely, when the 6S protocol is best, $\tilde{K}=-1$.

In Fig. \ref{optimalchannel}, from left to right, the protocols are operated at increasingly higher key rates.  Given a minimum key rate requirement, we compare the protocols which operate the best in bits per channel use. The main observation here is that the channel parameter space where the 6S protocol dominates shrinks for increasingly higher key rates and CV-QKD is at an advantage. It can also be seen that for higher minimum key-rate requirements, only the Sqz-Hom protocol can operate (see red regions in the middle and right subfigures). However, the 6S protocol can be operated in an intermediate-noise regime at low-key rates where CV-QKD cannot (left and centre subfigure). In Fig. \ref{ratioofnoise}, we plot the ratio of maximum noise tolerance of the Sqz-Hom protocol with the BB84 and the Six-State protocol for various $K_0$. Here it can be seen that DV-QKD performs better for low SKR requirements and at low transmittance values. 

Our results indicate that common QKD protocols are far from the upper bound secret key capacity in a thermal-loss channel. We also find that the NSqz-Hom protocol has the best excess noise tolerance in very noisy channels in agreement with Ref. \cite{Pirandola_2018} but we find that the BB84 and 6S protocols perform better in an intermediate noise regime. However, we note that Ref. \cite{usenko} arrives at different conclusions regarding the noise tolerance of DV-QKD compared to CV-QKD. The difference can be explained by the model for the thermal-loss channel in DV-QKD in Ref. \cite{usenko} which assumes that the signal and noise photons can be distinguished (see Appendix \ref{appendixH} for more detail).




\subsection{With phase noise $\sigma^2_\theta>0$}
In the following section, we quantify the performance of the 6S and Sqz-Hom (with optimized modulation variance $V_A$) protocols in the combined thermal-loss and phase noise channel. We note that we assume the preparation noise of the Sqz-Hom attributed to the squeezing angle $\phi$ is zero i.e. the squeezed states are perfectly amplitude or phase squeezed. First, consider the maximum tolerable thermal noise given by:

\begin{equation}
\begin{split}
    &N_{\text{Th}}^\text{(Max)} \\
    &=\text{arg} N_{\text{Th}}
    \begin{cases}
    &0, \qquad \text{if } K(0,\sigma_\theta^2,D)<K_0. \\
    &K(N_{\text{Th}},\sigma_\theta^2,D)=K_0, \text{ otherwise. }
\end{cases}
   \end{split}
\end{equation}
In other words, the maximum tolerable noise if the key rate is less than $K_0$ at $N_{\text{Th}}=0$ is $0$. Otherwise, the maximum tolerable noise is $N_{\text{Th}}$ when the key rate falls to $K_0$.

In Fig. \ref{phasenoise} a), we plot the following quantity:
\begin{equation}
\begin{split}    &\tilde{N}_{\text{Th}}^{\text{CV:DV}}(\sigma_\theta^2,D)
\\
&=\frac{N_{\text{Th,Sqz-Hom}}^\text{(Max)}-N_{\text{Th,6S}}^\text{(Max)}}{\text{Max}[N_{\text{Th,Sqz-Hom}}^\text{(Max)},N_{\text{Th,6S}}^\text{(Max)}]},
\end{split}
    \label{nmaxcv}
\end{equation} which is the difference between the maximum tolerable thermal noise of the Sqz-Hom and the 6S protocols for a given phase noise $\sigma_\theta^2$ and distance $D$ to achieve a key rate $K_0$. Highlighted in the figure is the green contour where both protocols tolerate the same amount of thermal noise, i.e. $\tilde{N}^{\text{CV:DV}}_{\text{Th}}=0$. For low key rates, it can be seen that the squeezed-state protocol tolerates more thermal noise than the 6S protocol in short channels and when $\sigma_\theta^2<10^{-3}$. The Sqz-Hom protocol also tolerates more thermal noise than the 6S protocol at longer distances. In this red region, the 6S protocol tolerates zero thermal noise, whereas the Sqz-Hom protocol tolerates some thermal noise. For higher key-rate requirements, although the region of noise tolerance shrinks for both protocols, the Sqz-Hom tolerates proportionally more thermal noise across the phase noise versus distance parameter space.   

Next, we consider the maximum distance or maximum tolerable loss given by:

\begin{equation}
\begin{split}
    &D^\text{(Max)} \\
    &=\text{arg} D
    \begin{cases}
    &0, \qquad \text{if } K(N_{\text{Th}},\sigma_\theta^2,0)<K_0. \\
    &K(N_{\text{Th}},\sigma_\theta^2,D)=K_0, \text{ otherwise. }
\end{cases}
   \end{split}
\end{equation}

In other words, the maximum distance if the key rate is less than $K_0$ at $D=0$ is $0$. Otherwise, the maximum distance is $D$ when the key rate falls to $K_0$. In Fig. \ref{phasenoise} b), we plot the following quantity:

\begin{equation}
\begin{split}
    &\tilde{L}^{\text{CV:DV}}(\sigma_\theta^2,N_{\text{Th}})=\tilde{D}^{\text{CV:DV}} (\sigma_\theta^2,N_{\text{Th}})\\
    &=\frac{D_{\text{Sqz-Hom}}^\text{(Max)}-D_{\text{6S}}^\text{(Max)}}{\text{Max}[D_{\text{Sqz-Hom}}^\text{(Max)},D_{\text{6S}}^\text{(Max)}]},
\end{split}
\label{maxloss}
\end{equation} which is the difference between the maximum distance of the Sqz-Hom and the 6S protocols for a given phase noise $\sigma_\theta^2$ and thermal noise $N_{\text{Th}}$ to achieve a key rate $K_0$. The Sqz-Hom protocol can tolerate more loss than the 6S protocol at thermal noise between $10^{-2}$ and $0.9$, and phase noise $\sigma_\theta^2<10^{-3}$. As found in \cite{usenko}, at a small region of high thermal noise, the 6S protocol tolerates more loss than the Sqz-Hom protocol. At higher key-rate requirements, the Sqz-Hom protocol can tolerate more loss compared to the 6S protocol. In fact, it can tolerate as much as $\sigma_\theta^2=0.05$ for a $K>10^{-3}$ key-rate requirement and $N_{\text{Th}}<10^{-3}$ to perform at a longer distance than the 6S protocol. 

From these results, we can conclude that for low key-rate requirements, the 6S protocol clearly dominates a larger region of parameters. However, for high key-rate requirements, the Sqz-Hom protocol dominates most of the parameter space for phase noise less than a phase noise of $\sigma^2_{\theta}<10^{-3}$.

\begin{center}
\begin{table}[h!]
\begin{tabular}{ |c|c|c| } 
\multicolumn{2}{c}{CV-QKD} \\ 
\hline
Reference & $\sigma_\theta^2$ \\
\hline
B. Qi et al. \cite{qi2015} & $4 \times 10^{-2} $ \\ 
\hline
T. Wang et al. \cite{Wang:18} & $1.2 \times 10^{-3} $ 
\\ 
\hline
H. Wang et al. \cite{wang20} & $\le 7.0 \times 10^{-3}$ \\ 
\hline
\multirow{2}{11em}{H.-M. Chin et al. \cite{Chin:2021} \& A. A. Hajomer et al. \cite{Hajomer:22}} & \multirow{2}{8em}{$\text{   }\le 1.0 \times 10^{-3}$}  \\ 
& \\ \hline
Y. Zhang et al. \cite{bestcvfibre} & $7.4 \times 10^{-5}$ \\ 
\hline 
\multicolumn{2}{c}{DV-QKD} \\ 
\hline
Reference & $\sigma_\theta^2$ \\
\hline
A. Boaron et al. \cite{bestqkd} & $7.2 \times 10^{-2}$ \\  \hline
W. Li et al. \cite{pan2023} & $2.2 \times 10^{-2}$ \\ 
\hline
\end{tabular}
\caption{Residual phase noise of locally generated local oscillator Gaussian modulated CV-QKD schemes in the first table and phase noise due to timing jitter in DV-QKD schemes in the second table. With the exception of Ref.  \cite{qi2015}, \cite{Wang:18}  \& \cite{bestcvfibre}, the phase noise is upper bounded from the total excess noise. }
\label{phasenoisetable}
\end{table}
\end{center}

As a comparison, experimental values for the phase noise in CV-QKD protocols are shown in Table. \ref{phasenoisetable}. These are also shown in Fig. \ref{phasenoise} a) along with current state-of-the-art DV-QKD protocols \cite{bestqkd} \& \cite{pan2023} (converted to equivalent distance in standard fibre). For DV-QKD implementations, we convert the time jitter Full Width at Half Maximum (FWHM) $\Delta t_{\text{FWHM}}$) to the phase noise using the following equation:
\begin{equation}
    \sigma^2_\theta=\frac{(2\pi \Delta t_ {\text{FWHM}})^2}{(2 \sqrt{2\ln{2}} \Delta t)^2},
\end{equation} where $\Delta t$ is the timing between pulses. The fundamental limitation of timing jitter is from the uncertainty in the arrival time of the photon. It is usually quoted as a FWHM which is defined as “the difference between the two values of the independent variable at which the dependent variable is equal to half of its maximum value”. This equation converts FWHM to a Gaussian width \cite{gwidth} and then to a phase noise (in radians squared). The timing between pulses in both experiments is inversely proportional to the repetition rate $\Delta t=1/f$.





\section*{Discussion}
\label{conclude}

We discuss our results with less-than-ideal experimental setups of QKD protocols. In optical fibre, the current distance record for DV-QKD is $421~\text{km}$ in ultralow-loss (ULL) fibre ($0.17~\text{dB/km}$) corresponding to $71.9~\text{dB}$ loss \cite{bestqkd}. A secret key rate of $0.25~\text{bps}$ or equivalently $K=10^{-10}~\text{bits per channel use}$ was obtained using superconducting single-photon detectors at a repetition rate of $2.5~\text{GHz}$. Most recently, a high key rate of $K=4.4 \times 10^{-2}$ was demonstrated in $10~\text{km}$ of standard optical fiber for DV-QKD \cite{pan2023}. We plot these experimental points normalized to standard optical fiber loss ($0.2~\text{ dB/km}$) in Fig. \ref{phasenoise} a). Based on these results, for this similar key-rate requirement $K_0=10^{-3}$, CV-QKD would, in theory, be able to achieve the same high key rate and tolerate more noise if the same levels of phase noise are maintained as in \cite{Wang:18} \& \cite{wang20}. Additionally, CV-QKD can extend up to $150~\text{ km}$ as opposed to DV-QKD which cannot tolerate noise beyond $125~\text{ km}$. In the rightmost subfigure in Fig. \ref{phasenoise} b), it can be seen that CV-QKD can tolerate more loss than DV-QKD for a large parameter region given a higher key rate requirement.  


Nonetheless, in terms of distance, DV-QKD is currently leading the benchmark for QKD with the world record for CV-QKD being more than half the distance in ULL fibre at $202.81~\text{km}$ (or $32.4~\text{dB}$) using the GG02 protocol where a key rate of $K\approx 10^{-6}$ was achieved \cite{bestcvfibre}. On the other hand, the apparent advantage of CV-QKD is in the efficient encoding of keys per symbol and the faster generation and detection of coherent (or squeezed) states with a much larger block size. Post-processing codes at low signal-to-noise ratio were a bottleneck in CV-QKD until it was recently shown that Raptor-like LPDC codes can maintain a high key extraction rate and high reconciliation efficiency, paving the way for practical and deployable CV-QKD \cite{recon}.

We have also focused mainly on the squeezed-state protocol. Despite renewed interest in the squeezed-state protocol due to its robustness to noise \cite{usenkouni, Derkach_2020,neda2022}, the difficulty of modulating and generating stable squeezed coherent states remains. However, entanglement-based versions have been demonstrated \cite{madsen}, and sources of highly entangled TMSV states are a promising pathway toward realizing the squeezed-state protocol \cite{wangy}.      

{ Furthermore, one of the limitations of CV-QKD is currently the maintaining of phase reference using a local oscillator (LO), which needs to be practically solved without compromising unconditional security, in a real-world setting outside of the laboratory \cite{bestcvfibre}. It can be seen from our results, that although CV-QKD performs well in a high-thermal noise regime, the introduction of phase noise destroys this advantage. For CV-QKD to maintain this advantage, the amount of phase noise must be less than $\sigma^2_\theta<10^{-3}$. However, we also find that CV-QKD performs best for high minimum key-rate requirements where it can tolerate more thermal noise at longer distances than DV-QKD. The physical reason behind this is that in CV-QKD, more symbols can be sent that will result in a shared key. Conversely, DV-QKD is limited to single photons.

Based on these results, we speculate that the consistent historical performance of DV-QKD protocols is mainly due to robustness to phase noise, which plagues CV-QKD protocols that rely on encoding information in phase as well as amplitude. However, with increasingly more robust carrier phase compensation schemes based on machine learning as in Ref. \cite{Chin:2021, Hajomer:22}, phase noise may no longer be a limiting factor in CV-QKD. 

}


Although current upper bounds on the secret key capacity can serve as a benchmark for QKD protocols, no QKD protocol is currently known to saturate these bounds in the thermal-loss channel. In our analysis, we have only considered point-to-point QKD, but fundamental bounds on the secret key capacity using repeaters exist \cite{repeaterbound}. Recently, it was shown that an entanglement purification scheme could saturate the repeater bounds in a pure-loss channel \cite{matt2022} but this remains an open problem in the thermal-loss channel. We note that energy-constrained upper bounds in the thermal-loss channel have been recently determined, that would be comparable in energy to common DV-QKD protocols \cite{mwilde}. Additionally, identifying the optimal QKD protocol for the phase diffusion channel is a task for future research.

\section{Conclusion}
In this work, we compared DV-QKD and CV-QKD protocols on equal grounds in a thermal-loss channel and we assumed ideal sources and detector performances. We developed analytical formulas for the QBER of the BB84 and six-state protocols in a thermal-loss channel. We introduced the minimum key rate as a metric for QKD performance. We found the squeezed-state protocol dominates most of the channel parameter regimes when there is no phase noise, except for an intermediate-noise regime where the six-state protocol can tolerate more loss and surpasses the lower bound to the secret key capacity. With the addition of phase noise, we find that the overall landscape of the DV-QKD and CV-QKD comparison becomes more complex. Finally, we find DV-QKD is largely unaffected by phase noise, whilst CV-QKD is sensitive but performs better below a threshold phase noise only recently reached in experiments.   


\bibliographystyle{unsrtnat}
\bibliography{sample}

\providecommand{\noopsort}[1]{}\providecommand{\singleletter}[1]{#1}%
\begin{thebibliography}{56}
\providecommand{\natexlab}[1]{#1}
\providecommand{\url}[1]{\texttt{#1}}
\expandafter\ifx\csname urlstyle\endcsname\relax
  \providecommand{\doi}[1]{doi: #1}\else
  \providecommand{\doi}{doi: \begingroup \urlstyle{rm}\Url}\fi

\bibitem[Hosseinidehaj et~al.(2019)Hosseinidehaj, Babar, Malaney, Ng, and Hanzo]{nedapredict}
Nedasadat Hosseinidehaj, Zunaira Babar, Robert Malaney, Soon~Xin Ng, and Lajos Hanzo.
\newblock Satellite-based continuous-variable quantum communications: State-of-the-art and a predictive outlook.
\newblock \emph{IEEE Communications Surveys Tutorials}, 21\penalty0 (1):\penalty0 881--919, 2019.
\newblock \doi{10.1109/COMST.2018.2864557}.

\bibitem[Bennett and Brassard(2014)]{BENNETT20147}
Charles~H. Bennett and Gilles Brassard.
\newblock Quantum cryptography: Public key distribution and coin tossing.
\newblock \emph{Theoretical Computer Science}, 560:\penalty0 7--11, 2014.
\newblock ISSN 0304-3975.
\newblock \doi{https://doi.org/10.1016/j.tcs.2014.05.025}.
\newblock URL \url{https://www.sciencedirect.com/science/article/pii/S0304397514004241}.
\newblock Theoretical Aspects of Quantum Cryptography – celebrating 30 years of BB84.

\bibitem[Su(2020)]{simplebb84}
Hong-Yi Su.
\newblock Simple analysis of security of the {BB84} quantum key distribution protocol.
\newblock \emph{Quantum Information Processing}, 19\penalty0 (6):\penalty0 169, 2020.
\newblock \doi{10.1007/s11128-020-02663-z}.
\newblock URL \url{https://doi.org/10.1007/s11128-020-02663-z}.

\bibitem[Ralph(1999)]{timsqueeze}
T.~C. Ralph.
\newblock Continuous variable quantum cryptography.
\newblock \emph{Phys. Rev. A}, 61:\penalty0 010303, Dec 1999.
\newblock \doi{10.1103/PhysRevA.61.010303}.
\newblock URL \url{https://link.aps.org/doi/10.1103/PhysRevA.61.010303}.

\bibitem[Ralph(2000)]{ralph2000}
T.~C. Ralph.
\newblock Security of continuous-variable quantum cryptography.
\newblock \emph{Phys. Rev. A}, 62:\penalty0 062306, Nov 2000.
\newblock \doi{10.1103/PhysRevA.62.062306}.
\newblock URL \url{https://link.aps.org/doi/10.1103/PhysRevA.62.062306}.

\bibitem[Hillery(2000)]{hillery2000}
Mark Hillery.
\newblock Quantum cryptography with squeezed states.
\newblock \emph{Phys. Rev. A}, 61:\penalty0 022309, Jan 2000.
\newblock \doi{10.1103/PhysRevA.61.022309}.
\newblock URL \url{https://link.aps.org/doi/10.1103/PhysRevA.61.022309}.

\bibitem[Cerf et~al.(2001)Cerf, L\'evy, and Assche]{assche2001}
N.~J. Cerf, M.~L\'evy, and G.~Van Assche.
\newblock Quantum distribution of gaussian keys using squeezed states.
\newblock \emph{Phys. Rev. A}, 63:\penalty0 052311, Apr 2001.
\newblock \doi{10.1103/PhysRevA.63.052311}.
\newblock URL \url{https://link.aps.org/doi/10.1103/PhysRevA.63.052311}.

\bibitem[Grosshans and Grangier(2002)]{gross1}
Fr\'ed\'eric Grosshans and Philippe Grangier.
\newblock Continuous variable quantum cryptography using coherent states.
\newblock \emph{Phys. Rev. Lett.}, 88:\penalty0 057902, Jan 2002.
\newblock \doi{10.1103/PhysRevLett.88.057902}.
\newblock URL \url{https://link.aps.org/doi/10.1103/PhysRevLett.88.057902}.

\bibitem[Grosshans et~al.(2003)Grosshans, Van~Assche, Wenger, Brouri, Cerf, and Grangier]{gross2}
Fr{\'e}d{\'e}ric Grosshans, Gilles Van~Assche, J{\'e}r{\^o}me Wenger, Rosa Brouri, Nicolas~J. Cerf, and Philippe Grangier.
\newblock Quantum key distribution using gaussian-modulated coherent states.
\newblock \emph{Nature}, 421\penalty0 (6920):\penalty0 238--241, 2003.
\newblock \doi{10.1038/nature01289}.
\newblock URL \url{https://doi.org/10.1038/nature01289}.

\bibitem[Weedbrook et~al.(2004)Weedbrook, Lance, Bowen, Symul, Ralph, and Lam]{cvhet}
Christian Weedbrook, Andrew~M. Lance, Warwick~P. Bowen, Thomas Symul, Timothy~C. Ralph, and Ping~Koy Lam.
\newblock Quantum cryptography without switching.
\newblock \emph{Phys. Rev. Lett.}, 93:\penalty0 170504, Oct 2004.
\newblock \doi{10.1103/PhysRevLett.93.170504}.
\newblock URL \url{https://link.aps.org/doi/10.1103/PhysRevLett.93.170504}.

\bibitem[Silberhorn et~al.(2002)Silberhorn, Ralph, L\"utkenhaus, and Leuchs]{silberhorn}
Ch. Silberhorn, T.~C. Ralph, N.~L\"utkenhaus, and G.~Leuchs.
\newblock Continuous variable quantum cryptography: Beating the 3 db loss limit.
\newblock \emph{Phys. Rev. Lett.}, 89:\penalty0 167901, Sep 2002.
\newblock \doi{10.1103/PhysRevLett.89.167901}.
\newblock URL \url{https://link.aps.org/doi/10.1103/PhysRevLett.89.167901}.

\bibitem[Wang et~al.(2019)Wang, Guo, Wang, Liu, and Li]{Wang:19}
Xuyang Wang, Siyou Guo, Pu~Wang, Wenyuan Liu, and Yongmin Li.
\newblock Realistic rate-distance limit of continuous-variable quantum key distribution.
\newblock \emph{Opt. Express}, 27\penalty0 (9):\penalty0 13372--13386, Apr 2019.
\newblock \doi{10.1364/OE.27.013372}.
\newblock URL \url{http://www.opticsexpress.org/abstract.cfm?URI=oe-27-9-13372}.

\bibitem[Pirandola et~al.(2020)Pirandola, Andersen, Banchi, Berta, Bunandar, Colbeck, Englund, Gehring, Lupo, Ottaviani, Pereira, Razavi, Shaari, Tomamichel, Usenko, Vallone, Villoresi, and Wallden]{Pirandola20}
S.~Pirandola, U.~L. Andersen, L.~Banchi, M.~Berta, D.~Bunandar, R.~Colbeck, D.~Englund, T.~Gehring, C.~Lupo, C.~Ottaviani, J.~L. Pereira, M.~Razavi, J.~Shamsul Shaari, M.~Tomamichel, V.~C. Usenko, G.~Vallone, P.~Villoresi, and P.~Wallden.
\newblock Advances in quantum cryptography.
\newblock \emph{Adv. Opt. Photon.}, 12\penalty0 (4):\penalty0 1012--1236, Dec 2020.
\newblock \doi{10.1364/AOP.361502}.
\newblock URL \url{http://aop.osa.org/abstract.cfm?URI=aop-12-4-1012}.

\bibitem[Pirandola et~al.(2015{\natexlab{a}})Pirandola, Ottaviani, Spedalieri, Weedbrook, Braunstein, Lloyd, Gehring, Jacobsen, and Andersen]{piran2}
Stefano Pirandola, Carlo Ottaviani, Gaetana Spedalieri, Christian Weedbrook, Samuel~L. Braunstein, Seth Lloyd, Tobias Gehring, Christian~S. Jacobsen, and Ulrik~L. Andersen.
\newblock High-rate measurement-device-independent quantum cryptography.
\newblock \emph{Nature Photonics}, 9\penalty0 (6):\penalty0 397--402, 2015{\natexlab{a}}.
\newblock \doi{10.1038/nphoton.2015.83}.
\newblock URL \url{https://doi.org/10.1038/nphoton.2015.83}.

\bibitem[Xu et~al.(2015)Xu, Curty, Qi, Qian, and Lo]{xu}
Feihu Xu, Marcos Curty, Bing Qi, Li~Qian, and Hoi-Kwong Lo.
\newblock Discrete and continuous variables for measurement-device-independent quantum cryptography.
\newblock \emph{Nature Photonics}, 9\penalty0 (12):\penalty0 772--773, 2015.
\newblock \doi{10.1038/nphoton.2015.206}.
\newblock URL \url{https://doi.org/10.1038/nphoton.2015.206}.

\bibitem[Pirandola et~al.(2015{\natexlab{b}})Pirandola, Ottaviani, Spedalieri, Weedbrook, Braunstein, Lloyd, Gehring, Jacobsen, and Andersen]{Pirandola2015B}
Stefano Pirandola, Carlo Ottaviani, Gaetana Spedalieri, Christian Weedbrook, Samuel~L. Braunstein, Seth Lloyd, Tobias Gehring, Christian~S. Jacobsen, and Ulrik~L. Andersen.
\newblock Reply to 'discrete and continuous variables for measurement-device-independent quantum cryptography'.
\newblock \emph{Nature Photonics}, 9\penalty0 (12):\penalty0 773--775, Dec 2015{\natexlab{b}}.
\newblock ISSN 1749-4893.
\newblock \doi{10.1038/nphoton.2015.207}.
\newblock URL \url{https://doi.org/10.1038/nphoton.2015.207}.

\bibitem[Lasota et~al.(2017)Lasota, Filip, and Usenko]{usenko}
Miko\l{}aj Lasota, Radim Filip, and Vladyslav~C. Usenko.
\newblock Robustness of quantum key distribution with discrete and continuous variables to channel noise.
\newblock \emph{Phys. Rev. A}, 95:\penalty0 062312, Jun 2017.
\newblock \doi{10.1103/PhysRevA.95.062312}.
\newblock URL \url{https://link.aps.org/doi/10.1103/PhysRevA.95.062312}.

\bibitem[Er-long et~al.(2005)Er-long, Zheng-fu, Shun-sheng, Tao, Da-sheng, and Guang-can]{erlong}
Miao Er-long, Han Zheng-fu, Gong Shun-sheng, Zhang Tao, Diao Da-sheng, and Guo Guang-can.
\newblock Background noise of satellite-to-ground quantum key distribution.
\newblock \emph{New Journal of Physics}, 7\penalty0 (1):\penalty0 215, oct 2005.
\newblock \doi{10.1088/1367-2630/7/1/215}.
\newblock URL \url{https://dx.doi.org/10.1088/1367-2630/7/1/215}.

\bibitem[Diamanti et~al.(2016)Diamanti, Lo, Qi, and Yuan]{eleni2016}
Eleni Diamanti, Hoi-Kwong Lo, Bing Qi, and Zhiliang Yuan.
\newblock Practical challenges in quantum key distribution.
\newblock \emph{npj Quantum Information}, 2\penalty0 (1):\penalty0 16025, 2016.
\newblock \doi{10.1038/npjqi.2016.25}.
\newblock URL \url{https://doi.org/10.1038/npjqi.2016.25}.

\bibitem[Chen et~al.(2021)Chen, Zhang, Chen, Cai, Liao, Zhang, Chen, Yin, Ren, Chen, Han, Yu, Liang, Zhou, Yuan, Zhao, Wang, Jiang, Zhang, Liu, Li, Shen, Cao, Lu, Shu, Wang, Li, Liu, Xu, Wang, Peng, and Pan]{pan2021}
Yu-Ao Chen, Qiang Zhang, Teng-Yun Chen, Wen-Qi Cai, Sheng-Kai Liao, Jun Zhang, Kai Chen, Juan Yin, Ji-Gang Ren, Zhu Chen, Sheng-Long Han, Qing Yu, Ken Liang, Fei Zhou, Xiao Yuan, Mei-Sheng Zhao, Tian-Yin Wang, Xiao Jiang, Liang Zhang, Wei-Yue Liu, Yang Li, Qi~Shen, Yuan Cao, Chao-Yang Lu, Rong Shu, Jian-Yu Wang, Li~Li, Nai-Le Liu, Feihu Xu, Xiang-Bin Wang, Cheng-Zhi Peng, and Jian-Wei Pan.
\newblock An integrated space-to-ground quantum communication network over 4,600 kilometres.
\newblock \emph{Nature}, 589\penalty0 (7841):\penalty0 214--219, 2021.
\newblock \doi{10.1038/s41586-020-03093-8}.
\newblock URL \url{https://doi.org/10.1038/s41586-020-03093-8}.

\bibitem[Shor and Preskill(2000)]{shor2000}
Peter~W. Shor and John Preskill.
\newblock Simple proof of security of the {BB84} quantum key distribution protocol.
\newblock \emph{Phys. Rev. Lett.}, 85:\penalty0 441--444, Jul 2000.
\newblock \doi{10.1103/PhysRevLett.85.441}.
\newblock URL \url{https://link.aps.org/doi/10.1103/PhysRevLett.85.441}.

\bibitem[Renner(2008)]{renner2008}
Renato Renner.
\newblock Security of quantum key distribution.
\newblock \emph{International Journal of Quantum Information}, 06\penalty0 (01):\penalty0 1--127, 2008.
\newblock \doi{10.1142/S0219749908003256}.
\newblock URL \url{https://doi.org/10.1142/S0219749908003256}.

\bibitem[Murta et~al.(2020)Murta, Rozp\ifmmode~\mbox{\k{e}}\else \k{e}\fi{}dek, Ribeiro, Elkouss, and Wehner]{murta2020}
Gl\'aucia Murta, Filip Rozp\ifmmode~\mbox{\k{e}}\else \k{e}\fi{}dek, J\'er\'emy Ribeiro, David Elkouss, and Stephanie Wehner.
\newblock Key rates for quantum key distribution protocols with asymmetric noise.
\newblock \emph{Phys. Rev. A}, 101:\penalty0 062321, Jun 2020.
\newblock \doi{10.1103/PhysRevA.101.062321}.
\newblock URL \url{https://link.aps.org/doi/10.1103/PhysRevA.101.062321}.

\bibitem[Nielsen and Chuang(2010)]{nielsen}
Michael~A. Nielsen and Isaac~L. Chuang.
\newblock Quantum computation and quantum information: 10th anniversary edition.
\newblock 2010.

\bibitem[Renner et~al.(2005)Renner, Gisin, and Kraus]{BB84noise}
Renato Renner, Nicolas Gisin, and Barbara Kraus.
\newblock Information-theoretic security proof for quantum-key-distribution protocols.
\newblock \emph{Phys. Rev. A}, 72:\penalty0 012332, Jul 2005.
\newblock \doi{10.1103/PhysRevA.72.012332}.
\newblock URL \url{https://link.aps.org/doi/10.1103/PhysRevA.72.012332}.

\bibitem[Sa\'nchez(2007)]{raulsanchez}
R.~Garcia-Patron Sa\'nchez.
\newblock Quantum information with optical continuous variables: from {Bell} tests to key distribution.
\newblock 2007.

\bibitem[Wolf et~al.(2006)Wolf, Giedke, and Cirac]{wolf}
Michael~M. Wolf, Geza Giedke, and J.~Ignacio Cirac.
\newblock Extremality of gaussian quantum states.
\newblock \emph{Phys. Rev. Lett.}, 96:\penalty0 080502, Mar 2006.
\newblock \doi{10.1103/PhysRevLett.96.080502}.
\newblock URL \url{https://link.aps.org/doi/10.1103/PhysRevLett.96.080502}.

\bibitem[Garc\'{\i}a-Patr\'on and Cerf(2006)]{patron}
Ra\'ul Garc\'{\i}a-Patr\'on and Nicolas~J. Cerf.
\newblock Unconditional optimality of gaussian attacks against continuous-variable quantum key distribution.
\newblock \emph{Phys. Rev. Lett.}, 97:\penalty0 190503, Nov 2006.
\newblock \doi{10.1103/PhysRevLett.97.190503}.
\newblock URL \url{https://link.aps.org/doi/10.1103/PhysRevLett.97.190503}.

\bibitem[Navascu\'es et~al.(2006)Navascu\'es, Grosshans, and Ac\'{\i}n]{nava}
Miguel Navascu\'es, Fr\'ed\'eric Grosshans, and Antonio Ac\'{\i}n.
\newblock Optimality of gaussian attacks in continuous-variable quantum cryptography.
\newblock \emph{Phys. Rev. Lett.}, 97:\penalty0 190502, Nov 2006.
\newblock \doi{10.1103/PhysRevLett.97.190502}.
\newblock URL \url{https://link.aps.org/doi/10.1103/PhysRevLett.97.190502}.

\bibitem[Laudenbach et~al.(2018)Laudenbach, Pacher, Fung, Poppe, Peev, Schrenk, Hentschel, Walther, and Hübel]{laudenbach}
Fabian Laudenbach, Christoph Pacher, Chi-Hang~Fred Fung, Andreas Poppe, Momtchil Peev, Bernhard Schrenk, Michael Hentschel, Philip Walther, and Hannes Hübel.
\newblock Continuous-variable quantum key distribution with gaussian modulation—the theory of practical implementations.
\newblock \emph{Advanced Quantum Technologies}, 1\penalty0 (1):\penalty0 1800011, 2018.
\newblock \doi{https://doi.org/10.1002/qute.201800011}.

\bibitem[Garc\'{\i}a-Patr\'on and Cerf(2009)]{garcianoise}
Ra\'ul Garc\'{\i}a-Patr\'on and Nicolas~J. Cerf.
\newblock Continuous-variable quantum key distribution protocols over noisy channels.
\newblock \emph{Phys. Rev. Lett.}, 102:\penalty0 130501, Mar 2009.
\newblock \doi{10.1103/PhysRevLett.102.130501}.
\newblock URL \url{https://link.aps.org/doi/10.1103/PhysRevLett.102.130501}.

\bibitem[Lamoureux et~al.(2005)Lamoureux, Brainis, Cerf, Emplit, Haelterman, and Massar]{dvphasenoise}
L.-P. Lamoureux, E.~Brainis, N.~J. Cerf, Ph. Emplit, M.~Haelterman, and S.~Massar.
\newblock Experimental error filtration for quantum communication over highly noisy channels.
\newblock \emph{Phys. Rev. Lett.}, 94:\penalty0 230501, Jun 2005.
\newblock \doi{10.1103/PhysRevLett.94.230501}.
\newblock URL \url{https://link.aps.org/doi/10.1103/PhysRevLett.94.230501}.

\bibitem[Lami and Wilde(2023)]{wilde-bosonic}
Ludovico Lami and Mark~M. Wilde.
\newblock Exact solution for the quantum and private capacities of bosonic dephasing channels.
\newblock \emph{Nature Photonics}, 17\penalty0 (6):\penalty0 525--530, 2023.
\newblock \doi{10.1038/s41566-023-01190-4}.
\newblock URL \url{https://doi.org/10.1038/s41566-023-01190-4}.

\bibitem[Pirandola et~al.(2017)Pirandola, Laurenza, Ottaviani, and Banchi]{plob}
Stefano Pirandola, Riccardo Laurenza, Carlo Ottaviani, and Leonardo Banchi.
\newblock Fundamental limits of repeaterless quantum communications.
\newblock \emph{Nature Communications}, 8\penalty0 (1):\penalty0 15043, 2017.
\newblock \doi{10.1038/ncomms15043}.
\newblock URL \url{https://doi.org/10.1038/ncomms15043}.

\bibitem[Pirandola et~al.(2009)Pirandola, Garc\'{\i}a-Patr\'on, Braunstein, and Lloyd]{rci}
Stefano Pirandola, Raul Garc\'{\i}a-Patr\'on, Samuel~L. Braunstein, and Seth Lloyd.
\newblock Direct and reverse secret-key capacities of a quantum channel.
\newblock \emph{Phys. Rev. Lett.}, 102:\penalty0 050503, Feb 2009.
\newblock \doi{10.1103/PhysRevLett.102.050503}.
\newblock URL \url{https://link.aps.org/doi/10.1103/PhysRevLett.102.050503}.

\bibitem[Vahlbruch et~al.(2016)Vahlbruch, Mehmet, Danzmann, and Schnabel]{squeezed15}
Henning Vahlbruch, Moritz Mehmet, Karsten Danzmann, and Roman Schnabel.
\newblock Detection of 15 db squeezed states of light and their application for the absolute calibration of photoelectric quantum efficiency.
\newblock \emph{Phys. Rev. Lett.}, 117:\penalty0 110801, Sep 2016.
\newblock \doi{10.1103/PhysRevLett.117.110801}.
\newblock URL \url{https://link.aps.org/doi/10.1103/PhysRevLett.117.110801}.

\bibitem[Pirandola et~al.(2018)Pirandola, Braunstein, Laurenza, Ottaviani, Cope, Spedalieri, and Banchi]{Pirandola_2018}
Stefano Pirandola, Samuel~L Braunstein, Riccardo Laurenza, Carlo Ottaviani, Thomas P~W Cope, Gaetana Spedalieri, and Leonardo Banchi.
\newblock Theory of channel simulation and bounds for private communication.
\newblock \emph{Quantum Science and Technology}, 3\penalty0 (3):\penalty0 035009, may 2018.
\newblock \doi{10.1088/2058-9565/aac394}.
\newblock URL \url{https://doi.org/10.1088/2058-9565/aac394}.

\bibitem[Qi et~al.(2015)Qi, Lougovski, Pooser, Grice, and Bobrek]{qi2015}
Bing Qi, Pavel Lougovski, Raphael Pooser, Warren Grice, and Miljko Bobrek.
\newblock Generating the local oscillator ``locally'' in continuous-variable quantum key distribution based on coherent detection.
\newblock \emph{Phys. Rev. X}, 5:\penalty0 041009, Oct 2015.
\newblock \doi{10.1103/PhysRevX.5.041009}.
\newblock URL \url{https://link.aps.org/doi/10.1103/PhysRevX.5.041009}.

\bibitem[Wang et~al.(2018)Wang, Huang, Zhou, Liu, Ma, Wang, and Zeng]{Wang:18}
Tao Wang, Peng Huang, Yingming Zhou, Weiqi Liu, Hongxin Ma, Shiyu Wang, and Guihua Zeng.
\newblock High key rate continuous-variable quantum key distribution with a real local oscillator.
\newblock \emph{Opt. Express}, 26\penalty0 (3):\penalty0 2794--2806, Feb 2018.
\newblock \doi{10.1364/OE.26.002794}.
\newblock URL \url{https://opg.optica.org/oe/abstract.cfm?URI=oe-26-3-2794}.

\bibitem[Wang et~al.(2020)Wang, Pi, Huang, Li, Shao, Yang, Liu, Zhang, Zhang, and Xu]{wang20}
Heng Wang, Yaodi Pi, Wei Huang, Yang Li, Yun Shao, Jie Yang, Jinlu Liu, Chenlin Zhang, Yichen Zhang, and Bingjie Xu.
\newblock High-speed gaussian-modulated continuous-variable quantum key distribution with a local local oscillator based on pilot-tone-assisted phase compensation.
\newblock \emph{Opt. Express}, 28\penalty0 (22):\penalty0 32882--32893, Oct 2020.
\newblock \doi{10.1364/OE.404611}.
\newblock URL \url{https://opg.optica.org/oe/abstract.cfm?URI=oe-28-22-32882}.

\bibitem[Chin et~al.(2021)Chin, Jain, Zibar, Andersen, and Gehring]{Chin:2021}
Hou-Man Chin, Nitin Jain, Darko Zibar, Ulrik~L. Andersen, and Tobias Gehring.
\newblock Machine learning aided carrier recovery in continuous-variable quantum key distribution.
\newblock \emph{npj Quantum Information}, 7\penalty0 (1):\penalty0 20, 2021.
\newblock \doi{10.1038/s41534-021-00361-x}.
\newblock URL \url{https://doi.org/10.1038/s41534-021-00361-x}.

\bibitem[Hajomer et~al.(2022)Hajomer, Mani, Jain, Chin, Andersen, and Gehring]{Hajomer:22}
Adnan~A.E. Hajomer, Hossein Mani, Nitin Jain, Hou-Man Chin, Ulrik~L. Andersen, and Tobias Gehring.
\newblock Continuous-variable quantum key distribution over 60 km optical fiber with real local oscillator.
\newblock In \emph{European Conference on Optical Communication (ECOC) 2022}, page Th1G.5. Optica Publishing Group, 2022.
\newblock URL \url{https://opg.optica.org/abstract.cfm?URI=ECEOC-2022-Th1G.5}.

\bibitem[Zhang et~al.(2020)Zhang, Chen, Pirandola, Wang, Zhou, Chu, Zhao, Xu, Yu, and Guo]{bestcvfibre}
Yichen Zhang, Ziyang Chen, Stefano Pirandola, Xiangyu Wang, Chao Zhou, Binjie Chu, Yijia Zhao, Bingjie Xu, Song Yu, and Hong Guo.
\newblock Long-distance continuous-variable quantum key distribution over 202.81 km of fiber.
\newblock \emph{Phys. Rev. Lett.}, 125:\penalty0 010502, Jun 2020.
\newblock \doi{10.1103/PhysRevLett.125.010502}.
\newblock URL \url{https://link.aps.org/doi/10.1103/PhysRevLett.125.010502}.

\bibitem[Boaron et~al.(2018)Boaron, Boso, Rusca, Vulliez, Autebert, Caloz, Perrenoud, Gras, Bussi\`eres, Li, Nolan, Martin, and Zbinden]{bestqkd}
Alberto Boaron, Gianluca Boso, Davide Rusca, C\'edric Vulliez, Claire Autebert, Misael Caloz, Matthieu Perrenoud, Ga\"etan Gras, F\'elix Bussi\`eres, Ming-Jun Li, Daniel Nolan, Anthony Martin, and Hugo Zbinden.
\newblock Secure quantum key distribution over 421 km of optical fiber.
\newblock \emph{Phys. Rev. Lett.}, 121:\penalty0 190502, Nov 2018.
\newblock \doi{10.1103/PhysRevLett.121.190502}.
\newblock URL \url{https://link.aps.org/doi/10.1103/PhysRevLett.121.190502}.

\bibitem[Li et~al.(2023)Li, Zhang, Tan, Lu, Liao, Huang, Li, Wang, Mao, Yan, Li, Liu, Zhang, Peng, You, Xu, and Pan]{pan2023}
Wei Li, Likang Zhang, Hao Tan, Yichen Lu, Sheng-Kai Liao, Jia Huang, Hao Li, Zhen Wang, Hao-Kun Mao, Bingze Yan, Qiong Li, Yang Liu, Qiang Zhang, Cheng-Zhi Peng, Lixing You, Feihu Xu, and Jian-Wei Pan.
\newblock High-rate quantum key distribution exceeding 110 mb s--1.
\newblock \emph{Nature Photonics}, 2023.
\newblock \doi{10.1038/s41566-023-01166-4}.
\newblock URL \url{https://doi.org/10.1038/s41566-023-01166-4}.

\bibitem[Caloz et~al.(2018)Caloz, Perrenoud, Autebert, Korzh, Weiss, Schönenberger, Warburton, Zbinden, and Bussières]{gwidth}
Misael Caloz, Matthieu Perrenoud, Claire Autebert, Boris Korzh, Markus Weiss, Christian Schönenberger, Richard~J. Warburton, Hugo Zbinden, and Félix Bussières.
\newblock {High-detection efficiency and low-timing jitter with amorphous superconducting nanowire single-photon detectors}.
\newblock \emph{Applied Physics Letters}, 112\penalty0 (6), 02 2018.
\newblock ISSN 0003-6951.
\newblock \doi{10.1063/1.5010102}.
\newblock URL \url{https://doi.org/10.1063/1.5010102}.
\newblock 061103.

\bibitem[Zhou et~al.(2021)Zhou, Wang, Zhang, Yu, Chen, and Guo]{recon}
Chao Zhou, XiangYu Wang, ZhiGuo Zhang, Song Yu, ZiYang Chen, and Hong Guo.
\newblock Rate compatible reconciliation for continuous-variable quantum key distribution using raptor-like ldpc codes.
\newblock \emph{Science China Physics, Mechanics {\&} Astronomy}, 64\penalty0 (6):\penalty0 260311, Apr 2021.
\newblock ISSN 1869-1927.
\newblock \doi{10.1007/s11433-021-1688-4}.
\newblock URL \url{https://doi.org/10.1007/s11433-021-1688-4}.

\bibitem[Usenko(2018)]{usenkouni}
Vladyslav~C. Usenko.
\newblock Unidimensional continuous-variable quantum key distribution using squeezed states.
\newblock \emph{Phys. Rev. A}, 98:\penalty0 032321, Sep 2018.
\newblock \doi{10.1103/PhysRevA.98.032321}.
\newblock URL \url{https://link.aps.org/doi/10.1103/PhysRevA.98.032321}.

\bibitem[Derkach et~al.(2020)Derkach, Usenko, and Filip]{Derkach_2020}
Ivan Derkach, Vladyslav~C Usenko, and Radim Filip.
\newblock Squeezing-enhanced quantum key distribution over atmospheric channels.
\newblock \emph{New Journal of Physics}, 22\penalty0 (5):\penalty0 053006, may 2020.
\newblock \doi{10.1088/1367-2630/ab7f8f}.
\newblock URL \url{https://dx.doi.org/10.1088/1367-2630/ab7f8f}.

\bibitem[Hosseinidehaj et~al.(2022)Hosseinidehaj, Winnel, and Ralph]{neda2022}
Nedasadat Hosseinidehaj, Matthew~S. Winnel, and Timothy~C. Ralph.
\newblock Simple and loss-tolerant free-space quantum key distribution using a squeezed laser.
\newblock \emph{Phys. Rev. A}, 105:\penalty0 032602, Mar 2022.
\newblock \doi{10.1103/PhysRevA.105.032602}.
\newblock URL \url{https://link.aps.org/doi/10.1103/PhysRevA.105.032602}.

\bibitem[Madsen et~al.(2012)Madsen, Usenko, Lassen, Filip, and Andersen]{madsen}
Lars~S. Madsen, Vladyslav~C. Usenko, Mikael Lassen, Radim Filip, and Ulrik~L. Andersen.
\newblock Continuous variable quantum key distribution with modulated entangled states.
\newblock \emph{Nature Communications}, 3\penalty0 (1):\penalty0 1083, 2012.
\newblock \doi{10.1038/ncomms2097}.
\newblock URL \url{https://doi.org/10.1038/ncomms2097}.

\bibitem[Wang et~al.(2021)Wang, Zhang, Li, Tian, and Zheng]{wangy}
Yajun Wang, Wenhui Zhang, Ruixin Li, Long Tian, and Yaohui Zheng.
\newblock Generation of $-$10.7 $\text{dB}$ unbiased entangled states of light.
\newblock \emph{Applied Physics Letters}, 118\penalty0 (13):\penalty0 134001, 2021.
\newblock \doi{10.1063/5.0041289}.
\newblock URL \url{https://doi.org/10.1063/5.0041289}.

\bibitem[Pirandola(2019)]{repeaterbound}
Stefano Pirandola.
\newblock End-to-end capacities of a quantum communication network.
\newblock \emph{Communications Physics}, 2\penalty0 (1):\penalty0 51, 2019.
\newblock \doi{10.1038/s42005-019-0147-3}.
\newblock URL \url{https://doi.org/10.1038/s42005-019-0147-3}.

\bibitem[Winnel et~al.(2022)Winnel, Guanzon, Hosseinidehaj, and Ralph]{matt2022}
Matthew~S. Winnel, Joshua~J. Guanzon, Nedasadat Hosseinidehaj, and Timothy~C. Ralph.
\newblock Achieving the ultimate end-to-end rates of lossy quantum communication networks.
\newblock \emph{npj Quantum Information}, 8\penalty0 (1):\penalty0 129, 2022.
\newblock \doi{10.1038/s41534-022-00641-0}.
\newblock URL \url{https://doi.org/10.1038/s41534-022-00641-0}.

\bibitem[Davis et~al.(2018)Davis, Shirokov, and Wilde]{mwilde}
Noah Davis, Maksim~E. Shirokov, and Mark~M. Wilde.
\newblock Energy-constrained two-way assisted private and quantum capacities of quantum channels.
\newblock \emph{Phys. Rev. A}, 97:\penalty0 062310, Jun 2018.
\newblock \doi{10.1103/PhysRevA.97.062310}.
\newblock URL \url{https://link.aps.org/doi/10.1103/PhysRevA.97.062310}.

\bibitem[Marie and All\'eaume(2017)]{marie2017}
Adrien Marie and Romain All\'eaume.
\newblock Self-coherent phase reference sharing for continuous-variable quantum key distribution.
\newblock \emph{Phys. Rev. A}, 95:\penalty0 012316, Jan 2017.
\newblock \doi{10.1103/PhysRevA.95.012316}.
\newblock URL \url{https://link.aps.org/doi/10.1103/PhysRevA.95.012316}.

\end{thebibliography}

\section*{Acknowledgements}
We wish to acknowledge Spyros Tserkis, and Matthew Winnel for their valuable discussions. We thank the reviewers for their valuable comments that led to the improvement of this article. This research was supported by the Australian Research Council (ARC) under the Centre of Excellence for Quantum Computation and Communication Technology (Grant No. CE110001027).

\section*{Author contributions statement}

S. P. K. wrote the paper, and produced the majority of calculations and results. P. J. G. produced the DV phase-noise model in Sec 2, DV thermal-loss-to-depolarising derivation (Appendix B), and helped develop the squeezed-state phase-noise model. A. W. produced some of the calculations in the Appendix,  S. M. A. and P. K. L. conceived the main idea and proofread the manuscript.  All authors reviewed the manuscript. 

\begin{appendix}
\renewcommand{\thesubsection}{\Alph{subsection}}
\section*{Appendices}
\appendix
\subsection{Quantum Bit Error Rate (QBER)}
\label{appendixa}
To calculate $Q_Z$, we consider the probability of a bit-flip if Alice sends a logical $\bf{0}$ (i.e. $\ket{1}_{a_1} \ket{0}_{a_2}$) and Bob detects a logical $\bf{1}$ (i.e. simultaneously detects $\ket{0}_{b_1}$ and $\ket{1}_{b_2}$) given by,
\begin{equation}
\begin{split}
&P_{Z,\bf{0}\rightarrow\bf{1}}=P_{Z,\ket{1}_{a_1} \rightarrow \ket{0}_{b_1}} P_{Z,\ket{0}_{a_2} \rightarrow \ket{1}_{b_2}} \\
&=\text{Tr}( \hat{U}_{BS}(\eta) (\ket{1}_{a_1} \bra{1}_{a_1} \otimes \hat{\rho}_{\text{Th}} ) \hat{U}_{BS}^\dagger(\eta) \ket{0}_{b_1} \bra{0}_{b_1}) \\
&\times \text{Tr}( \hat{U}_{BS}(\eta) (\ket{0}_{a_2} \bra{0}_{a_2} \otimes \hat{\rho}_{\text{Th}} ) \hat{U}_{BS}^\dagger(\eta) \ket{1}_{b_2} \bra{1}_{b_2}),
\end{split} 
\end{equation} where $\hat{\rho}_{\text{Th}}=\sum^{\infty}_{n=0} [{N_{\text{Th}}^n}/{(N_{\text{Th}}+1)^{n+1}}] \ket{n}\bra{n}$ is the thermal state with average thermal photon number $N_{\text{Th}}$ and $\hat{U}_{BS}(\eta)$ is the unitary beamsplitter transformation mixing the thermal environment and the state sent by Alice. If Alice prepares a logical $\bf{1}$ and Bob measures $\bf{0}$, the probability is
\begin{equation}
\begin{split}
&P_{Z,\bf{1}\rightarrow\bf{0}}=P_{Z,\ket{0}_{a_1} \rightarrow \ket{1}_{b_1}} P_{Z,\ket{1}_{a_2} \rightarrow \ket{0}_{b_2}} \\
&=\text{Tr} ( \hat{U}_{BS}(\eta) (\ket{0}_{a_1} \bra{0}_{a_1} \otimes \hat{\rho}_{\text{Th}} ) \hat{U}_{BS}^\dagger(\eta) \ket{1}_{b_1} \bra{1}_{b_1}) \\
&\times \text{Tr} ( \hat{U}_{BS}(\eta) (\ket{1}_{a_2} \bra{1}_{a_2} \otimes \hat{\rho}_{\text{Th}} ) \hat{U}_{BS}^\dagger(\eta) \ket{0}_{b_2} \bra{0}_{b_2}),
\end{split} 
\end{equation} and since we assume the channels are symmetric, $P_{Z,\bf{1}\rightarrow\bf{0}}=P_{Z,\bf{0}\rightarrow\bf{1}}$. The total un-normalized probability of a bit-flip is $2P_{Z,\bf{0}\rightarrow\bf{1}}$.

Bob only accepts the correct bits and the flipped bits. Therefore, we normalize by considering the total probability Bob only detects the logical bits in the $Z$-basis. Hence
\begin{equation}
Q_Z=\frac{P_{Z,\bf{0}\rightarrow\bf{1}}}{P_{Z,\bf{0}\rightarrow\bf{1}}+P_{Z,\bf{0}\rightarrow\bf{0}}}=\frac{P_{Z,\bf{1}\rightarrow\bf{0}}}{P_{Z,\bf{1}\rightarrow\bf{0}}+P_{Z,\bf{1}\rightarrow\bf{1}}},
\label{qz}
\end{equation} where $P_{Z,\bf{0}\rightarrow\bf{0}}=P_{Z,\ket{1}_{a_1} \rightarrow \ket{1}_{b_1}} P_{Z,\ket{0}_{a_2} \rightarrow \ket{0}_{b_2}}$ and $P_{Z,\bf{1}\rightarrow\bf{1}}=P_{Z,\ket{0}_{a_1} \rightarrow \ket{0}_{b_1}} P_{Z,\ket{1}_{a_2} \rightarrow \ket{1}_{b_2}}$ are the probabilities of Bob detecting the same bits that Alice sent after passing through the channel. These probabilities are given by
\begin{equation}
\begin{split}
&P_{Z,\bf{0}\rightarrow\bf{0}}=P_{Z,\bf{1}\rightarrow\bf{1}}\\
&=\text{Tr}( \hat{U}_{BS}(\eta) (\ket{1}_{a_1} \bra{1}_{a_1} \otimes \hat{\rho}_{\text{Th}} ) \hat{U}_{BS}^\dagger(\eta) \ket{1}_{b_1} \bra{1}_{b_1}) \\
&\times \text{Tr}( \hat{U}_{BS}(\eta) (\ket{0}_{a_2} \bra{0}_{a_2} \otimes \hat{\rho}_{\text{Th}} ) \hat{U}_{BS}^\dagger(\eta) \ket{0}_{b_2} \bra{0}_{b_2}).
\end{split}
\end{equation}
These probabilities are:
\begin{align}
P_{Z,\bf{0}\rightarrow\bf{1}}&= P_{Z,\bf{1}\rightarrow\bf{0}}= \frac{(1-\eta)^2(N_{\text{Th}}+N_{\text{Th}}^2)}{(1+(1-\eta)N_{\text{Th}})^4} \\
P_{Z,\bf{0}\rightarrow\bf{0}}&=P_{Z,\bf{1}\rightarrow\bf{1}}=\frac{\eta+(1-\eta)^2(N_{\text{Th}}+N_{\text{Th}}^2)}{(1+(1-\eta)N_{\text{Th}})^4}.
\end{align}
To calculate $Q_X$, we consider the bit-flips in the $X$ basis. In this case, the modes $a_1$ and $a_2$ are entangled because of the balanced beamsplitter (see Fig. \ref{rect} b)). Therefore, we consider the joint probability given by 
\begin{widetext}
\begin{equation}
\begin{split}
&P_{X,\bf{0}\rightarrow\bf{1}}\\
&=\text{Tr}[ \hat{U}_{50/50,b_1 b_2}\hat{U}_{BS,a_1}(\eta) \hat{U}_{BS,a_2}(\eta) ( \ket{-}_{a_1,a_2} \bra{-}_{a_1,a_2} \otimes \hat{\rho}_{a_1,\text{Th}}  \otimes \hat{\rho}_{a_2,\text{Th}}) \hat{U}_{BS,a_1}^\dagger(\eta)\hat{U}_{BS,a_2}^\dagger(\eta) \hat{U}_{50/50,b_1 b_2}^\dagger M_{\bf{1}} ], \\
&P_{X,\bf{1}\rightarrow\bf{0}}\\
&=\text{Tr}[ \hat{U}_{50/50,b_1 b_2 }\hat{U}_{BS,a_1}(\eta) \hat{U}_{BS,a_2}(\eta) ( \ket{+}_{a_1,a_2} \bra{+}_{a_1,a_2} \otimes \hat{\rho}_{a_1,\text{Th}}  \otimes \hat{\rho}_{a_2,\text{Th}}) \hat{U}_{BS,a_1}^\dagger(\eta)\hat{U}_{BS,a_2}^\dagger(\eta) \hat{U}_{50/50,b_1 b_2}^\dagger M_{\bf{0}} ],
\end{split}
\end{equation} 
\end{widetext}
where $\hat{U}_{50/50,b_1 b_2}$ is the second balanced beamsplitter unitary, $\ket{-}_{a_1,a_2} \bra{-}_{a_1,a_2} =R_{\pi,a_1} \ket{+}_{a_1,a_2} \bra{+}_{a_1,a_2} R^\dagger_{\pi,a_1}$ is obtained by applying a $\pi$-phase shifter to the state $\ket{+}$, $M_{\bf{1}} =\ket{0}_{b'_1}\bra{0}_{b'_1} \otimes \ket{1}_{b'_2} \bra{1}_{b'_2}$ is the logical $\bf{1}$ measurement outcome and $M_{\bf{0}} =\ket{1}_{b'_1}\bra{1}_{b'_1} \otimes \ket{0}_{b'_2} \bra{0}_{b'_2}$. Similar to above, we also renormalize to obtain the QBER,
\begin{equation}
Q_X=\frac{P_{X,\bf{0}\rightarrow\bf{1}}}{P_{X,\bf{0}\rightarrow\bf{1}}+P_{X,\bf{0}\rightarrow\bf{0}}}.
\label{qx}
\end{equation}
We find due to symmetry that the probabilities for the diagonal basis are the same as for the rectilinear basis and it follows that $Q_X=Q_Z$, simplifying the key rate equation. 

\subsection{Thermal-loss to depolarized state}
\label{depolar}
\begin{figure}[b]
  \centering
  \includegraphics[width=0.4\textwidth]{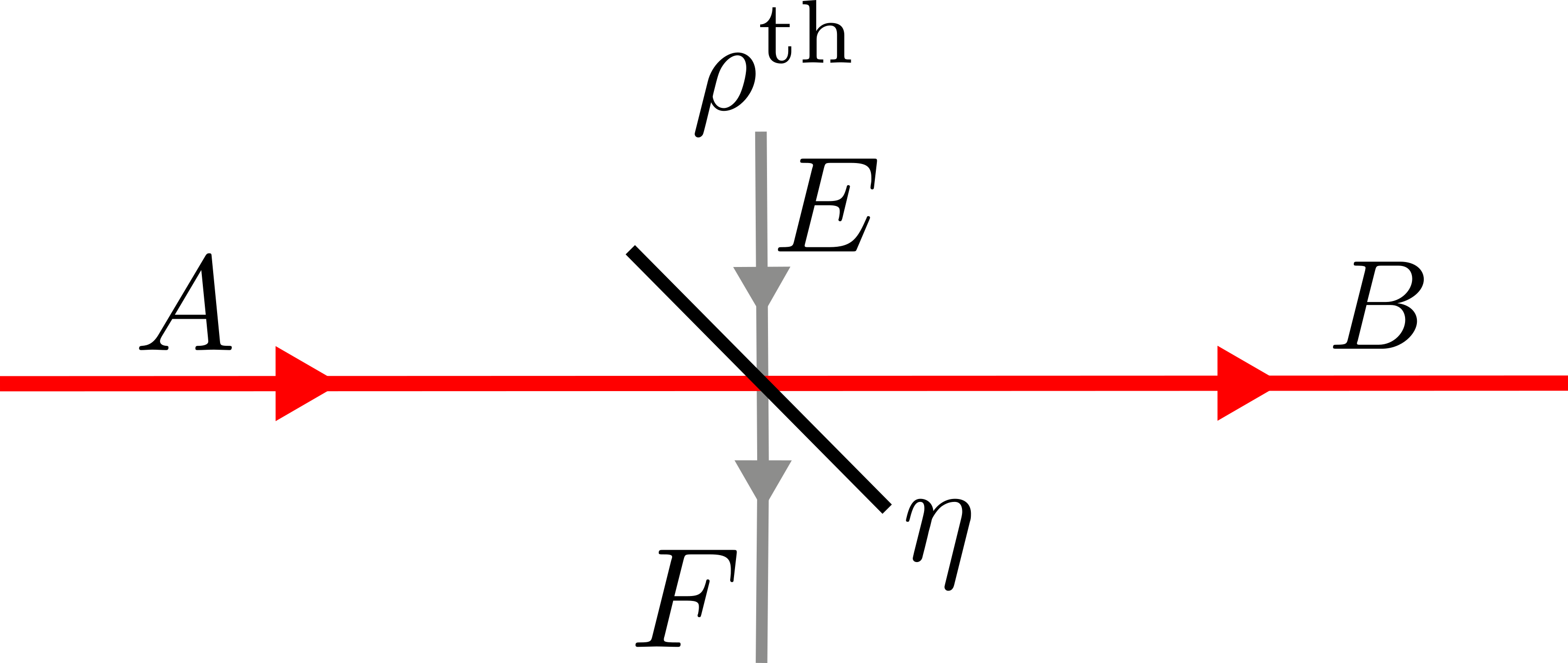}
  \caption{A thermal noise rail with modes labeled for Alice, Bob, and the environment.}
  \label{fig:thermal-rail}
\end{figure}
Using the model of thermal noise from the previous section we identify Alice's input mode $A$, Bob's output mode $B$, and the environmental input and output modes $E$ and $F$ (see Fig. \ref{fig:thermal-rail}), with corresponding creation and annihilation operators (lowercase). A photon-number (Fock) state of a bosonic mode may be expressed as $\ket{n} = \frac{(\ah\hc)^n}{\sqrt{n!}}\ket{0}$; in this representation, the action of the beamsplitter is given entirely by the transformation
\begin{equation}\label{eq:bs}
\begin{aligned}
  \ah \mapsto \sqrt{\eta}\,\bh + \sqrt{1-\eta}\,\fh \\[1ex]
  \eh \mapsto \sqrt{1-\eta}\,\bh - \sqrt{\eta}\,\fh.
\end{aligned}
\end{equation}
If the beamsplitter receives no photon from Alice and exactly $n$ photons from the environment, then under action \eqref{eq:bs} the combined input state $\ket{0,n}_{AE}$ transforms as
\begin{widetext}
\begin{align*}
  \ket{0,n}_{AE} = \frac{(\eh\hc)^n}{\sqrt{n!}}\ket{0,0}_{AE} &\longmapsto \frac{(\sqrt{1-\eta}\,\bh\hc - \sqrt{\eta}\,\fh\hc)^n}{\sqrt{n!}}\ket{0,0}_{BF} \\
  &= \frac{1}{\sqrt{n!}} \left(\sum_{k=0}^n \binom{n}{k}(\sqrt{1-\eta}\;\bh\hc)^{n-k}(-\sqrt{\eta}\;\fh\hc)^k\right)\ket{0,0} \\
  &= \frac{1}{\sqrt{n!}} \sum_{k=0}^n \frac{n!}{k!(n-k)!}(\sqrt{1-\eta})^{n-k}(-\sqrt{\eta})^k \sqrt{(n-k)!}\sqrt{k!}\ket{n-k,k} \\
  &= \sum_{k=0}^n \sqrt{\binom{n}{k}}(\sqrt{1-\eta})^{n-k}(-\sqrt{\eta})^k\ket{n-k,k}\\
  &:= \ket{\psi_n}
\end{align*}
\end{widetext}
which is a coherent superposition of Fock states, with $n$ total photons split across rails $B$ and $F$ according to a binomial distribution. If Alice instead sends a single photon we obtain
\begin{widetext}
\begin{align*}
  \ket{1,n}_{AE} = \ah\hc \ket{0,n}_{AE} &\longmapsto (\sqrt{\eta}\,\bh\hc + \sqrt{1-\eta}\,\fh\hc) \ket{\psi_n} \\
    &= \sqrt{\eta}\sum_{k=0}^n \sqrt{\binom{n}{k}} (-\sqrt{\eta})^k(\sqrt{1-\eta})^{n-k}\sqrt{n-k+1} \ket{n-k+1,k} \\
    &\qquad+\sqrt{1-\eta}\sum_{k=0}^n\sqrt{\binom{n}{k}} (-\sqrt{\eta})^k(\sqrt{1-\eta})^{n-k}\sqrt{k+1}\ket{n-k,k+1} \\
    &:= \ket{\phi_n}
\end{align*}
\end{widetext}
where the first term corresponds with Alice's photon reaching Bob, and the second with it escaping to the environment.

It follows that Alice's input can be considered a $2\times 2$ density matrix $\hat{\rho}^A$ with terms of form $\rho^A_{ij}\ketbra{\bi}{\bj}$. The collective input $AE$ to the beamsplitter system is therefore
\[\hat{\rho}^{\mathrm{in}} = \hat{\rho}^A \otimes \hat{\rho}^E = \sum_{i,j,\bn} \hat{\rho}^A_{ij} p_{\bn} \ketbra{\bi,\bn}{\bj,\bn}.\]

Since quantum channels are linear, the collective output $BF$ is determined by the action of the channel on each $\ketbra{\bi,\bn}{\bj,\bn}$ term (despite $\ketbra{\bi}{\bj}$ individually representing a nonphysical state whenever $i\neq j$). Since $\ket{\bi,\bn}$ represents an independent input to each beamsplitter, the output is a direct tensor product of the independent single-rail outputs derived above, i.e.
\begin{align*}
    \ket{\bi,\bn}_{AE} \longmapsto &\ket{\psi_{n_0}}_{B_0F_0}\ket{\psi_{n_1}}_{B_1F_1}\cdots\\
    &\ket{\phi_{n_0}}_{B_0 F_0}\cdots\ket{\psi_{n_{1}}}_{B_{n_{1}}F_{n_{1}}} := \ket{\win}_{BF}
    \end{align*}
where only rail $i$ has output $\ket{\phi_n}$, the  symbol $\omega$ was chosen for no particular reason. The Hermitian conjugate of Eq. \eqref{eq:bs} transforms the corresponding bra in the same way, giving $\bra{\bj,\bn} \longmapsto \bra{\omega_{\bj,\bn}}$ and hence $\ketbra{\bi,\bn}{\bj,\bn} \mapsto \outijn$. Bob's final state is
\begin{widetext}
\begin{equation}\label{eq:rhob}
\hat{\rho}^B = \Tr_F(\hat{\rho}^{\mathrm{out}}) = \Tr_F\left(\sum_{i,j,\bn} \rho^A_{ij} p_{\bn} \outijn\right) = \sum_{i,j} \rho^A_{ij} \sum_{\bn}p_{\bn} \Tr_F \outijn
\end{equation}
\end{widetext}
obtained by tracing over the environmental modes in the collective output $\hat{\rho}^{\mathrm{out}}$. 

We assume that Bob may perform a perfect photon-number-resolving (PNR) measurement in any desired basis, and that like Alice he is interested only in single-photon states $\ket{\bbeta} = \sum_i \beta_i \ket{\bi}$ and will discard all others. With perfect measurement, Bob's outcome probabilities are given by projection:
\begin{equation}\label{eq:prob}
  P(\ket{\bbeta}) = \expval{\hat{\rho}^B}{\bbeta} = \Tr(\ketbra{\bbeta}{\bbeta} \hat{\rho}^B)
\end{equation}
where only terms of form $\ketbra{\bi}{\bj}$ in Bob's state $\hat{\rho}^B$ contribute to this expression if $\ket{\bbeta}$ is a single-photon state. Like $\hat{\rho}^A$, Bob's state $\hat{\rho}^B$ may therefore be effectively considered a $2 \times 2$ matrix $\rho^B_{ij}$, which we now compute. 
Discarding terms which contain multiple photons in any single one of Bob's rails leaves 
\begin{widetext}
\begin{equation}
\begin{split}
  \ket{\psi_n} &\longrightarrow \ket{\psi_n'} = (-\sqrt{\eta})^{n-1}\left(-\sqrt{\eta}\ket{0,n} + \sqrt{n(1-\eta)}\ket{1,n-1}\right) \\
  \ket{\phi_n} &\longrightarrow \ket{\phi_n'} = (-\sqrt{\eta})^{n-1}\left([n-\eta n - \eta] \ket{1,n} - \sqrt{\eta(1-\eta)(n+1)}\ket{0,n+1}\right).
  \end{split}
\end{equation}
\end{widetext}
Next, to compute
\[\Tr_F \outijn = \sum_{\bn} \braket{\bn}{\win}\braket{\omega_{\bj,\bn}}{\bn}\]
we need only consider components of $\outijn$ with diagonal environmental mode $\ketbra{\bn}{\bn}$, as all others vanish. Discarding these nondiagonal terms in each of our single-rail outer products gives
\begin{widetext}
\begin{align}
  \ketbra{\psi_n'}{\psi_n'} &\longrightarrow \eta^{n-1}\bigl(\eta\ketbra{0,n}{0,n} + n(1-\eta)\ketbra{1,n-1}{1,n-1}\bigr)\label{eq:psipsi}\\[2ex]
  \ketbra{\phi_n'}{\phi_n'} &\longrightarrow \eta^{n-1}\bigl([n-\eta n - \eta]^2 \ketbra{1,n}{1,n} + \eta(1-\eta)(n+1)\ketbra{0,n+1}{0,n+1}\bigr)\label{eq:phiphi}\\[2ex]
  \ketbra{\phi_n'}{\psi_n'} &\longrightarrow -\eta^{n-1}\sqrt{\eta}\,[n-\eta n - \eta] \ketbra{1,n}{0,n}\label{eq:phipsi}\\[2ex]
  \ketbra{\psi_n'}{\phi_n'} &\longrightarrow -\eta^{n-1}\sqrt{\eta}\,[n-\eta n - \eta] \ketbra{0,n}{1,n}.\label{eq:psiphi}
\end{align}
\end{widetext}
We can decompose $\outijn$ in the collective Fock basis as a sum of terms corresponding with each different combination of photon numbers from Eqs. \eqref{eq:psipsi} and/or \eqref{eq:phiphi}. However, we keep only those terms with a photon in exactly one of Bob's modes; if $i \neq j$, terms \eqref{eq:phipsi} and \eqref{eq:psiphi} provide these photons (albeit in a different rail on each side of the outer product) and hence all other rails must be empty. After simultaneously tracing out the environment, this gives
\begin{widetext}
\[\Tr_F\outijn = \frac{1}{\eta}\left(\eta^{n_i}[n_i-\eta n_i - \eta]\right)\left(\eta^{n_j}[n_j-\eta n_j - \eta]\right)\left(\prod_{k\neq i,j} \eta^{n_k}\right) \ketbra{\bi}{\bj}.\]
\end{widetext}
If $i=j$, the photon is received either in the original rail $i$ or an erroneous rail $j \neq i$, giving
\begin{align*}
&\Tr_F\ketbra{\win}{\win}= \eta^{n_i-1}(n_i-\eta n_i - \eta) \left(\prod_{k\neq i} \eta^{n_k}\right) \ketbra{\bi}{\bi} \\
  &+ \sum_{j\neq i}(1-\eta)^2\,\eta^{n_i}(n_i+1)\,n_j \eta^{n_j-1} \left(\prod_{k\neq i,j} \eta^{n_k}\right) \ketbra{\bj}{\bj}.
\end{align*}
Returning to Eq. \eqref{eq:rhob}, we now sum over all $\bn$. This is done analytically, and can also be done with the aid of Mathematica. The resulting action of the channel is defined by
\begin{align*}
  \ketbra{\bi}{\bj}_A &\longmapsto \frac{\eta}{\gamma^{4}} \ketbra{\bi}{\bj}_B: \qquad i \neq j, \\[2ex]
  \ketbra{\bi}{\bi}_A &\longmapsto \frac{\eta}{\gamma^{4}} \ketbra{\bi}{\bi} + \frac{N_{\text{Th}}(1+N_{\text{Th}})(1-\eta)^2}{\gamma^{4}}\I
\end{align*}
where $\gamma = 1 + N_{\text{Th}} - N_{\text{Th}} \eta$ and $\I$ is the identity, i.e. $\I/2$ is the maximally-mixed state. Noting that $\Tr\hat{\rho}^A = \sum_i \rho^A_{ii} = 1$, we can thus express this as the qubit transformation $\hat{\rho}^A \to \hat{\rho}^B$ (see Eq. \eqref{eq:rhob}):
\begin{align}\label{eq:thermal-transformation}
\hat{\rho}^A &\longmapsto \frac{\eta}{\gamma^{4}}\hat{\rho}^A + \frac{N_{\text{Th}}(1+N_{\text{Th}})(1-\eta)^2}{\gamma^{4}}\left(\sum_i\hat{\rho}^A_{ii}\right)\I \\ 
&= \frac{\eta}{\gamma^{4}}\hat{\rho}^A + \frac{N_{\text{Th}}(1+N_{\text{Th}})(1-\eta)^2}{\gamma^{4}}\I.
\end{align}
The trace of this un-normalised output now represents the probability $P_s$ of successfully receiving a valid qubit:
\begin{equation}\label{eq:thermal-ps}
P_S = \Tr \hat{\rho}^B = \frac{\eta + 2\,N_{\text{Th}}(1+N_{\text{Th}})(1-\eta)^2}{\gamma^{4}}.
\end{equation}
Conditional on success, we obtain the normalised state
\begin{equation}
\begin{split}
\tilde{\hat{\rho}}^B := \frac{\hat{\rho}^B}{P_S} &= \frac{\eta}{\eta + 2\,N_{\text{Th}}(1+N_{\text{Th}})(1-\eta)^2} \hat{\rho}^A \\
&+ \frac{N_{\text{Th}}(1+N_{\text{Th}})(1-\eta)^2}{\eta + 2\,N_{\text{Th}}(1+N_{\text{Th}})(1-\eta)^2}\I.
\end{split}
\end{equation}
This represents a depolarizing channel \cite{nielsen}
\begin{equation}\label{eq:depolarising-channel}
  \hat{\rho} \to (1-\lambda)\hat{\rho} + \frac{\lambda}{2}\I
\end{equation}
with depolarizing parameter
\begin{equation}\label{eq:depolarising-parameter}
\lambda = \frac{2\,N_{\text{Th}}(1+N_{\text{Th}})(1-\eta)^2}{\eta + 2\,N_{\text{Th}}(1+N_{\text{Th}})(1-\eta)^2},
\end{equation}
which tends to 1 as $\eta \to 0$ or $N_{\text{Th}} \to \infty$, as expected. 

A property of the depolarizing channel is that the error rate is the same in all bases:
\begin{equation}\label{eq:thermal-qber}
Q = \frac{\lambda}{2} = \frac{N_{\text{Th}}(1+N_{\text{Th}})(1-\eta)^2}{\eta + 2 N_{\text{Th}}(1+N_{\text{Th}})(1-\eta)^2},
\end{equation}
which can be seen from Eq. \eqref{eq:depolarising-channel}. In this article, we only focus on the dual-rail case of $d=2$. It is left for future work to consider the high-dimensional QKD protocols in-depth.

\subsection{Fighting noise with noise squeezed state protocol}
\label{appendixwithnoise}
Introducing trusted Gaussian noise $\xi_{\text{B}}$ before Bob's detection modifies the mutual information:
\begin{equation}
I_{\text{AB}}^{\text{noise}}=\frac{1}{2}  \log_2{\Big( \frac{V_\text{B}+\xi_{\text{B}}}{V_{\text{B|A}}+\xi_{\text{B}}} \Big)},
\end{equation} 
The conditional entropy is:
\begin{equation}
    S(\text{E}|x_\text{B})=S(\text{BC}|x_\text{B})=G((\lambda_3-1)/2)+G((\lambda_4-1)/2),
\end{equation} where the symplectic eigenvalues are given by:
\begin{equation}
    \lambda^2_{3,4}=\frac{1}{2}[A\pm\sqrt{A^2-4B}],
\end{equation} where
\begin{equation}
    \begin{split}
        A&=\frac{1}{V_\text{B}+\xi_{\text{B}}}(V_\text{B}+V_\text{A} D+\xi_{\text{B}}\Delta), \\
        B&=\frac{D}{V_\text{B}+\xi_{\text{B}}}(V_\text{A}+\xi_{\text{B}} D),
    \end{split}
\end{equation} where for $\xi_{\text{B}}=0$ and $\xi_{\text{B}}=1$, we obtain the squeezed protocol with homodyne and heterodyne detection, respectively.

\subsection{Thermal-loss channel treatment in Ref. \cite{usenko}}

\label{appendixH}
\begin{figure}[t!]
\centering
\includegraphics[width=0.42\textwidth,trim=0mm 50mm 10mm 50mm, clip=true]{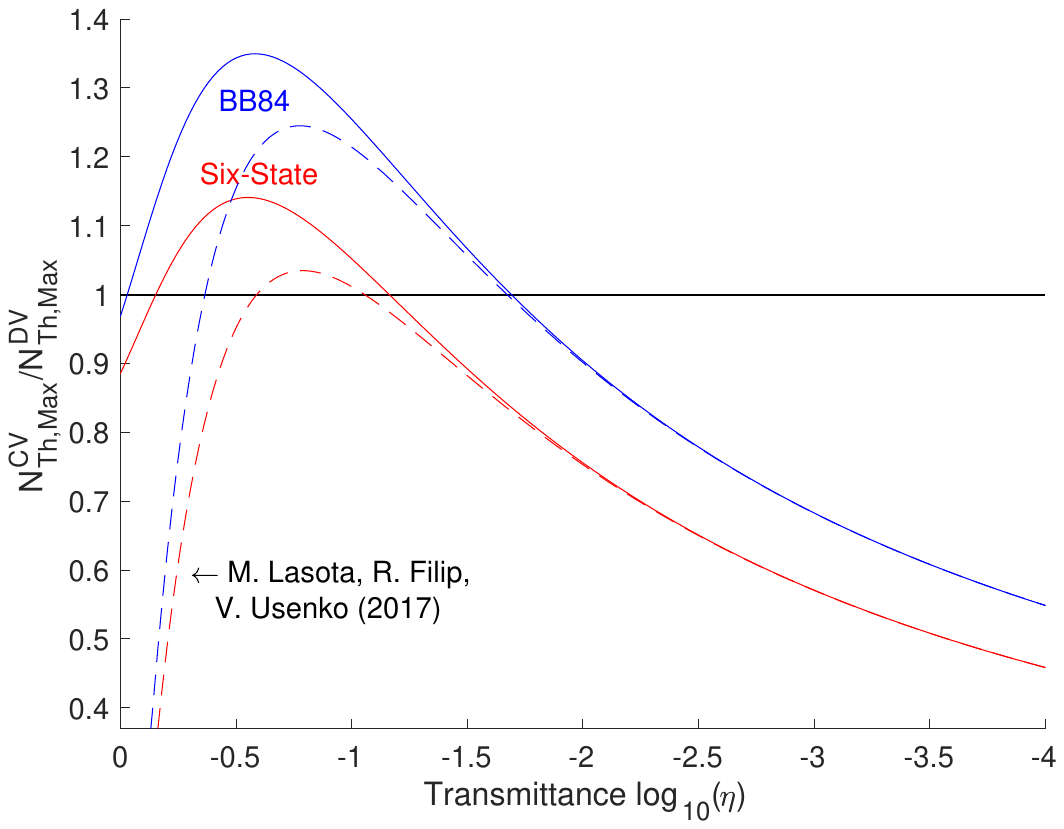}
\caption{Ratio of maximum noise tolerance of the squeezed state protocol with BB84 (blue) and Six-State (red) protocols. Shown are numerical results for $K_0$ of $10^{-10}$ (solid line), compared with the noise tolerance from Ref. \cite{usenko} (dashed lines). The protocol parameters of CV and DV are the same as those in Fig. \ref{lowthermal}.}
\label{usenkocomp}
\end{figure}

In Fig. \ref{usenkocomp}, we plot the numerical results of the noise tolerance ratio between CV and DV for comparison with Ref. \cite{usenko}. We find that the Sqz-Hom protocol tolerates thermal noise for a wider range of transmittance values for high secret key rate requirements. The numerical results from Ref. \cite{usenko} are different for $\eta=10^{-1}$ to $\eta=1$ which showed the ratio $N^{CV}_{\text{Th,Max}}/N^{DV}_{\text{Th,Max}}=0.4$ as $\eta \rightarrow 1$ from $\eta=10^{-0.5}=0.32$. Our results show a much less dramatic noise tolerance drop-off of the ratio compared to Ref. \cite{usenko}. The difference can be explained by the model for the thermal-loss channel in DV-QKD in Ref. \cite{usenko} which assumes that the signal and noise photons can be distinguished. The probabilities are calculated as a classical conditional probability rather than quantum as we will show below. In our model, we do not make this assumption, as the dual-rail protocol in the thermal-loss channel is completely controlled by Eve and consequently the QBER in our model is higher.
In this Appendix, we derive the equations that give the same QBER as in Ref. \cite{usenko}. The difference between our model in the main text and Ref. \cite{usenko} are \begin{enumerate}
    \item The treatment of the thermal noise model is different. Ref. \cite{usenko} treats the thermal-loss channel as a conditional photon-loss channel and a thermal noise channel.
    \item In their initial analysis they make use of ON/OFF detectors rather than photon-number resolving detectors. Nonetheless, these were shown to be equivalent for the noise tolerance.
\end{enumerate}
The first point of difference is the most important as this impacts the noise tolerance. 

With this model of noise, the accepted probability for the ON/OFF detector is given by \cite{usenko}:
\begin{equation}
    p_{\text{exp}}=\sum^{n_+=\infty}_{k=0} p_+(k,0)+\sum^{n_-=\infty}_{k=1} p_-(k,0)+\sum^{n_-=\infty}_{l=1} p_-(0,l),
\end{equation} where the first term represents the following:
\begin{equation}
    p_+(k,0)=[p_\eta(1,1) p_{\eta,N_{\text{Th}}}(0,k)] p_{\eta,N_{\text{Th}}}(0,0),
\end{equation} where $p_\eta(1,1)$ is the probability that a photon is not lost by the pure-loss channel multiplied by the probability $p_{\eta,N_{\text{Th}}}(0,k)$ that a noise photon is added in the RIGHT detector (inside of brackets) and multiplied by the probability $p_{\eta,N_{\text{Th}}}(0,0)$ no noise photon is added in the WRONG detector (outside of brackets). The middle term is:
\begin{equation}
    p_-(k,0)=[p_\eta(1,0) p_{\eta,N_{\text{Th}}}(0,k)] p_{\eta,N_{\text{Th}}}(0,0),
\end{equation} where $p_\eta(1,0)$ is the probability that a photon is lost by the pure-loss channel multiplied by the probability $p_{\eta,N_{\text{Th}}}(1,k)$ that a noise photon is added in the RIGHT detector and as before, multiplied by the probability $p_{\eta,N_{\text{Th}}}(0,0)$ no noise photon is added in the WRONG detector. Finally the last term is

\begin{equation}
    p_-(0,l)=[p_\eta(1,0) p_{\eta,N_{\text{Th}}}(0,0)] p_{\eta,N_{\text{Th}}}(0,l),
\end{equation} where a photon is lost by the pure-loss channel and no noise photon is added in the RIGHT detector and a noise photon is added in the WRONG detector.
Evidently this last term is the only term that contributes to the QBER:
\begin{equation}
    Q=\frac{\sum^{n=\infty}_{l=1} p_-(0,l)}{p_{\text{exp}}}.
\end{equation} 
For a photon number resolving detector (PNRD) the summations terminate at $n_+=0$ and $n_-=1$. Evaluating this expression for PNRDs and using the calculations from Appendix \ref{appendixa} that $p_{\eta,N_\text{Th}}(1,0)=(1-\eta) (N_\text{Th}+1)/\gamma^2$, $p_{\eta,N_\text{Th}}(0,1)=(1-\eta) N_\text{Th}/\gamma^2$, $p_{\eta,N_\text{Th}}(0,0)=1/\gamma$ and $p_{\eta,N_\text{Th}}(1,1)=(\eta+(1-\eta)^2 (N_\text{Th}^2+N_\text{Th}))/\gamma^3$ where it is reminded that $\gamma=1+(1-\eta) N_\text{Th}$. The probability of the bit-flip error is
\begin{equation}
    p_-(0,l)=\frac{N_\text{Th} (1-\eta)^2}{\gamma^3}.
\end{equation} 

Comparing this probability of a bit-flip with that of Appendix \ref{appendixa}:

\begin{equation}
\begin{split}
    P_{Z,\bf{0 \rightarrow 1}}&=\frac{\,N_{\text{Th}}(1+N_{\text{Th}})(1-\eta)^2}{\gamma^{4}}.
\end{split}
\end{equation}
Therefore, not being able to distinguish signal photons and noise photons increases the probability of a bit-flip error by the factor $(1+N_\text{Th})/(1+(1-\eta) N_\text{Th})>1$. The difference is that in the RIGHT detector the probability $p_\eta(1,0) p_{\eta,N_{\text{Th}}}(0,0)=(1-\eta)/\gamma\neq  p_{\eta,N_{\text{Th}}}(1,0)=(1-\eta)(N_\text{Th}+1)/\gamma^2$. The probability of success is $p_\text{exp}=\frac{\eta+N_{\text{Th}}(1-\eta)(2-\eta)}{\gamma^3}$.

Comparing the QBER of this ``distinguishable" approach:
\begin{equation}
    Q_\text{D}=\frac{N_\text{Th}(1-\eta)^2}{\eta+N_{\text{Th}}(1-\eta)(2-\eta)},
    \label{QBERD}
\end{equation} with calculations in Appendix \ref{appendixa} for indistinguishable noise and signal photons:

\begin{equation}
     Q_\text{I}= \frac{N_{\text{Th}}(1+N_{\text{Th}})(1-\eta)^2}{\eta + 2N_{\text{Th}}(1+N_{\text{Th}})(1-\eta)^2}.
\end{equation}
We remark that $Q_\text{D}$ for ON/OFF detectors are exactly the same as for PNRDs for this approach. We can evaluate the sums $\sum^{n=\infty}_{m=0} p_{\eta,N_\text{Th}}(0,m)=1$ and $\sum^{n=\infty}_{m=1} p_{\eta,N_\text{Th}}(0,m)=1-p_{\eta,N_\text{Th}}(0,0)=\frac{\gamma-1}{\gamma}$ leading to the probability of bit-flip errors $ \sum^{n=\infty}_{l=1} p_-(0,l)=N_\text{Th} (1-\eta)^2/\gamma^2$ and $p_\text{exp}=\frac{\eta+N_{\text{Th}}(1-\eta)(2-\eta)}{\gamma^2}$.

In Fig. \ref{usenkocomp}, we plot the ratio of the maximum noise tolerance of the SS-Hom protocol with the BB84 and 6S protocols, comparing the two approaches. As illustrated in Fig. \ref{usenkocomp}, distinguishable photons seem to tolerate more thermal noise than indistinguishable photons in the limit of $\eta \rightarrow 1$. Setting $Q_\text{D}=Q_\text{I}=0.1262$ for the highest QBER tolerance of the 6S protocol, the maximum tolerable noises are:

\begin{equation}
    N_\text{Th,D,Max}=\frac{631 \eta}{3738-8107\eta+4369 \eta^2},
    \label{told}
\end{equation} implies that at $\eta \rightarrow 0.8556$, the tolerable noise approaches $\infty$. For $\eta>0.8556$ the QBER never reaches $0.1262$ for any value of $N_\text{Th}$. And for indistinguishable:  

\begin{equation}
    N_\text{Th,I,Max}=\frac{\sqrt{1-\frac{2476 \eta}{1869}+\eta^2}}{2(1-\eta)}-\frac{1}{2},
    \label{Imax}
\end{equation} where we note that $N_\text{Th} < \eta/(1-\eta)$ for non-entanglement breaking channels. Clearly, Eq. (\ref{told}) violates this condition whereas Eq. (\ref{Imax}) does not. We note the difference with the results of Ref. \cite{usenko} which  determine the noise tolerance numerically. We believe that with high enough dimensions, this behaviour can be seen and will match the analytical expression for the QBER of Eq. (\ref{QBERD}).

In the high loss limit $\eta\ll 1$, the maximum tolerable noises are:
\begin{equation}
    N_\text{Th,D,Max}\approx \frac{0.1688 \eta}{1-2.1688 \eta},
\end{equation} and

\begin{equation}
    N_\text{Th,I,Max}\approx \frac{0.1688 \eta}{1-\eta}.
\end{equation} 

These results show that for $\eta\ll10^{-2}$ as seen in Fig. \ref{usenkocomp}, the ratio of CV to DV noise tolerances are both approaching the same limits i.e. $0.1688 \eta$. As loss increases and tolerable noise decreases, the ability to distinguish signal photons from noise photons is no longer an advantage.

\subsection{Squeezed-state protocol with phase noise}
\label{bestphasenoise}
In this section, we derive the covariance matrix of the squeezed-state protocol in the combined thermal-loss and phase noise channel. Consider the squeezed-state protocol where Alice's modulation variance is $V_{\text{sig}}=e^{2r}-e^{-2r}$. This state passes through a thermal-loss channel with $\eta$ and $N_{\text{Th}}$ followed by a phase noise channel with circular variance $V_\theta=\sigma_\theta^2$. Since Bob's average state is thermal, the phase noise does not affect the variance of Bob's data. Modeling the phase following the wrapped normal distribution as done for the DV case, Bob's received variance in the EB scheme is $V_B=\eta \mu+(1-\eta) (2 N_\text{Th}+1)$ where $\mu=V_\text{sig}+V_\text{sqz}=e^{2r}$. 

The inferred transmittance $\eta_{\text{I}}$ is found by first considering the Gaussian integrals of the correlations $\braket{X_A X_B | \theta }$ in the PM scheme for a fixed value of $\theta$:
\begin{equation}
\begin{split}
    &\braket{X_A X_B | \theta }\\
    &=\int \int dx_A dx_B f_{B|A} (x_A,x_B) f_A (x_A) x_A x_B \\
    &= V_\text{sig} \sqrt{\eta} \cos{\theta},
\end{split}
\end{equation} where for a fixed $\theta$, $x_A$ follows a Gaussian distribution (centered at zero) with modulation variance $V_\text{sig}$ and $x_B=\sqrt{\eta} x_A \cos{\theta} $ with variance $\Xi$ due to channel noise. The correlations are reduced as expected, due to the wrapping of the squeezed state around phase space, by a factor of $\cos{\theta}$. Next, we integrate over the wrapped normal distribution of $\theta$ which is also known as the circular mean as in Eq. (\ref{wrapped}). Therefore,
\begin{equation}
    \braket{X_A X_B}=\int d \theta V_\text{sig}\sqrt{\eta} f_{WN}(\theta) \cos{\theta}=V_\text{sig} \sqrt{\eta} \Bar{r},
\end{equation} where $\Bar{r}=e^{-\sigma_\theta^2/2}$. In the entanglement-based scheme, the correlations are $\braket{X_A X_B}_\text{EB}=\sqrt{\eta \Bar{r}^2 (\mu^2-1)}$. The inferred transmittance is therefore
\begin{equation}
    \eta_{\text{I}}=\eta e^{-\sigma_\theta^2}.
\end{equation}
The effective Gaussian channel is:
\begin{equation}
    V_B=\eta_{\text{I}} (\mu+\chi_I),
\end{equation} where $\chi_{\text{I}}$ is the inferred noise. Rearranging,
\begin{equation}
\begin{split}
    \chi_{\text{I}}&=\frac{V_B-\eta_I\mu}{\eta_\text{I}} \\
    &=\frac{(1-e^{-\sigma^2_\theta})\eta\mu+(1-\eta)(2 N_\text{Th}+1)}{\eta e^{-\sigma^2_\theta}}.
\end{split}
\end{equation}

We can also define the excess noise w.r.t. Alice due to phase noise $\xi_\theta$ by
\begin{equation}
    V_B=\eta_I \mu+(1-\eta)(2 N_\text{Th}+1)+\eta \xi_\theta,
\end{equation} implying that $\xi_\theta=(1-e^{-\sigma^2_\theta}) \mu$.

For the squeezed-state protocol, the combined thermal-loss and phase noise channel leads to the following covariance matrix in the EB scheme:
\begin{equation}
\begin{split}
&\gamma_{\rm AB}=
\begin{pmatrix}
a\mathbb{I} & c\sigma_{z} \\ c\sigma_{z} & b\mathbb{I}
\end{pmatrix}
 \\
&
=\begin{pmatrix}
\mu\mathbb{I} & \sqrt{\eta_\text{I}(\mu^2-1})\sigma_{z} \\ \sqrt{\eta_\text{I}(\mu^2-1})\sigma_{z} & (\eta_\text{I} (\mu+ \chi_\text{I})\mathbb{I}
\end{pmatrix} \\
&
=\begin{pmatrix}
\mu\mathbb{I} & \Bar{r}\sqrt{\eta (\mu^2-1})\sigma_{z} \\ \Bar{r}\sqrt{\eta(\mu^2-1})\sigma_{z} & (\eta \mu+(1-\eta) (2 N_\text{Th}+1))\mathbb{I}
\end{pmatrix},
\end{split}
\label{covariance3}
\end{equation} where the inferred transmittance $\eta_I=\eta \Bar{r}^2=\eta e^{-\sigma^2_\theta}$ and noise $\chi_I=\frac{(1-e^{-\sigma^2_\theta})\eta\mu+(1-\eta)(2 N_\text{Th}+1)}{\eta e^{-\sigma^2_\theta}}$.

The phase noise channel is a non-Gaussian channel. To estimate the channel parameters, Alice and Bob evaluate the covariance matrices between their measurement results and infer the effective transmittance ($\eta_I$) and effective noise ($\chi_I$) of an equivalent Gaussian channel. The phase noise reduces the effective transmittance, and increases the effective noise. 

We note that the phase noise model in Eq. (\ref{covariance3}) is different from previous treatments, for example, as in \cite{marie2017} for the GG02 protocol (see Appendix \ref{gg02phasenoise}) which only considers the effect of the phase noise on $\braket{\Delta X_B^2}$. The implicit assumption is that the estimate for $\eta$ (or $G$ defined in \cite{marie2017} as the intensity transmission of the channel) is unchanged and an excess noise $\xi_\theta$ is added by the phase noise channel. Without the attenuation in transmittance by the factor $\Bar{r}^2$, this leads to a model that is inaccurate in the high phase noise regime where the noise $\chi= \xi_\theta+\frac{(1-\eta)(2 N_\text{Th}+1)}{\eta}$ would be underestimated compared to $\chi_I$.  
\subsection{GG02 protocol phase noise}
\label{gg02phasenoise}
For the coherent state Gaussian modulated protocols, Alice prepares coherent states in the phase space with $\theta$ phase.

As in \cite{marie2017}, we consider the residual phase noise after estimating the phase. The quadratures after homodyne or heterodyne measurements are:

\begin{equation}
\begin{split}
 \begin{pmatrix}
x_m  \\
p_m  
\end{pmatrix}&=\sqrt{\frac{G}{2}}
    \begin{pmatrix}
\cos{\theta} & \sin{\theta} \\
-\sin{\theta} & \cos{\theta} 
\end{pmatrix}
\begin{pmatrix}
x_A+x_0  \\
p_A+p_0  
\end{pmatrix},
\end{split}
\label{xp}
\end{equation} where Alice sends a coherent state that follows a Gaussian distribution with $x_A \sim \mathcal{N}(0,V_x)$ and $p_A \sim \mathcal{N}(0,V_p)$ centered at $x_0=0$ and $p_0=0$ measured with a coherent detector with total intensity transmission $G$ of the channel.

Bob then estimates the phase with the estimator $\hat{\theta} \sim \mathcal{N}(\theta,V_\theta)$. Bob then sends his phase estimates to Alice who makes corrections and estimates Bob's measurements. The excess noise due to the phase noises would then be:
\begin{equation}
    \begin{split}
\xi_x&=\text{var}(x_m-\tilde{x}_m) \\
\xi_p&=\text{var}(p_m-\tilde{p}_m). 
    \end{split}
\end{equation} where $\text{var}$ is the variance, and $\tilde{x}_m$ and $\tilde{p}_m$ are the estimated quadratures as a function of the estimator $\hat{\theta}$. The excess noise depends on the remaining phase noise $\Theta=\theta-\hat{\theta}$ which we assume is a normally distributed variable $\Theta \sim \mathcal{N}(0,\sigma^2_{\Theta})$.
Then it is straightforward to calculate the excess noise:
\begin{equation}
    \begin{split}
\xi_x&=2 V_A (1-e^{-\tilde{\sigma}^2_{\Theta}/2}) \\
\xi_p&=2 V_A (1-e^{-\tilde{\sigma}^2_{ \Theta}/2}),
    \end{split}
\end{equation} where $\tilde{\sigma}^2_{ \Theta}=V_\theta$. For small phase noise $\tilde{\sigma}^2_{ \Theta}<0.1$, $\xi_x=\xi_p \approx \tilde{\sigma}^2_{ \Theta} V_A$.



\subsection{GG02 protocol with heterodyne detection}
\label{appendixgg02}
For heterodyne detection by Bob, the mutual information $I_{AB}$ in a thermal-loss channel is \cite{raulsanchez}
\begin{equation}
\begin{split}
    I_{\text{AB}}&=\log_2{\Bigg( \frac{V_\text{B}+1}{V_{\text{B}|\text{A}^M}+1} \Bigg)}\\
    &=\log_2{\Bigg( \frac{\eta V_\text{A}+(1-\eta)(2 N_{\text{Th}}+1)}{\eta+(1-\eta)(2 N_{\text{Th}}+1)} \Bigg)},
\end{split}
\label{mutual}
\end{equation} where $V_\text{B}$ is Bob's variance and $V_{\text{B}|\text{A}^M}=b-c^2/(a+1)$ is Bob's variance conditioned on Alice's heterodyne measurement. 
$S(\text{E|B})=S(\text{A}|x_B,p_D)$ is the information obtained by Eve conditioned on Bob's heterodyne measurement result $x_\text{B}$ and the auxiliary mode $p_\text{D}$ \cite{raulsanchez}. 
The covariance matrix of Alice after a projective measurement by Bob's heterodyne detection is
\begin{equation}
\gamma_\text{A}^{\text{out}}=\gamma_\text{A}-\sigma_{\text{AB}}(\gamma_\text{B}+\mathbb{I})^{-1} \sigma_{\text{AB}}^T,
\end{equation} where $\sigma_{\text{AB}}=c \sigma_Z$. The conditional Von Neumann entropy is
\begin{equation}
S(\text{A}|x_\text{B},p_\text{D})=G[(\lambda_3-1)/2],
\label{cond}
\end{equation} where the symplectic eigenvalue $\lambda_3$ is
\begin{equation}
\lambda_3=a-c^2/(b+1).
\label{lambda}
\end{equation} 

\subsection{Squeezing required for the Sqz-Hom protocol}
\label{bb84vssqueeze}
\begin{figure*}[b!]
\centering
\includegraphics[scale=0.4]{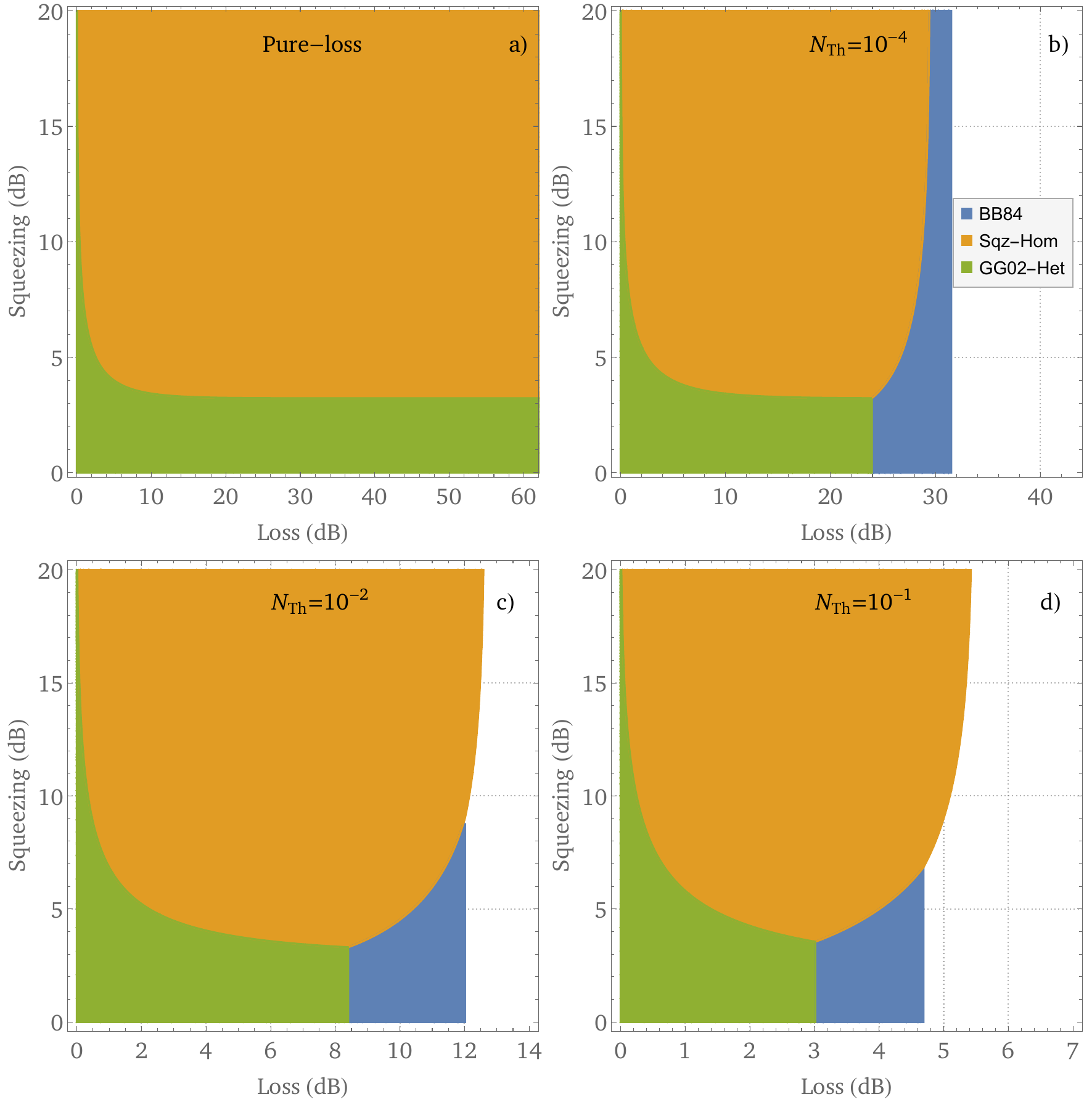}
\caption{Regions where QKD protocols give the highest secret key rate greater than $K=10^{-10}$ based on the amount of squeezing $V_{\text{sq}}$ prepared by Alice for the squeezed-state protocol with homodyne detection. In the unshaded regions, $K$ is less than $10^{-10}$ for all protocols. Comparison of the squeezed-state protocol with homodyne detection in a pure-loss channel based on the amount of squeezing prepared by Alice. Above $9 \text{ dB}$ of squeezing, the Sqz-Hom protocol performs better than GG02 and BB84 protocols.}
\label{squeezing}
\end{figure*}

In Fig. \ref{squeezing}, we compare the performance of the Sqz-Hom protocol for the amount of squeezing used to the BB84 protocol and GG02 with heterodyne (GG02) protocol. In a pure-loss channel (see Fig. \ref{squeezing} a)), Sqz-Hom protocols with more than $10 \text{ dB}$ of squeezing are sufficient to be equal to or better than the GG02 protocol for all loss parameters (where the key rate is greater than $K=10^{-10}$). However, for an intermediate-noise region (i.e. Fig. \ref{squeezing} b)), the BB84 protocol is robust at higher channel losses. We find that for very noisy thermal-loss channels shown in Fig. \ref{squeezing} c) and d), more than $9 \text{ dB}$ of squeezing is required to surpass BB84.

\end{appendix}
\end{document}